\documentclass[12pt]{article}

\usepackage{epsf, amssymb}
\usepackage{cite}
\usepackage{epsfig}
\usepackage{amsmath}
\usepackage{latexsym}
\usepackage{epsfig}
\usepackage[english]{babel}
\usepackage{graphicx,color}
\usepackage{url}

\usepackage[colorlinks=true, linkcolor=blue, bookmarks=true]{hyperref}

\makeatletter
\renewcommand\section{\@startsection {section}{1}{\z@}%
                               {-3.5ex \@plus -1ex \@minus -.2ex}%
                               {2.3ex \@plus.2ex}%
                               {\normalfont\large\bfseries}}
\renewcommand\subsection{\@startsection{subsection}{2}{\z@}%
                                 {-3.25ex\@plus -1ex \@minus -.2ex}%
                                 {1.5ex \@plus .2ex}%
                                 {\normalfont\bfseries}}
\makeatother

\def\IZ{\relax\ifmmode\mathchoice
{\hbox{\cmss Z\kern-.4em Z}}{\hbox{\cmss Z\kern-.4em Z}}
{\lower.9pt\hbox{\cmsss Z\kern-.4em Z}} {\lower1.2pt\hbox{\cmsss
Z\kern-.4em Z}}\else{\cmss Z\kern-.4em Z}\fi}
\def\IR{\relax{\rm I\kern-.18em R}}
\def\eg{{\it e.g.}}
\def\one{{\hbox{ 1\kern-.8mm l}}}

\def\tr{{\rm tr\,}}

\newlength{\bredde}
\def\slash#1{\settowidth{\bredde}{$#1$}\ifmmode\,\raisebox{.15ex}{/}
\hspace*{-\bredde} #1\else$\,\raisebox{.15ex}{/}\hspace*{-\bredde}
#1$\fi}

\newsavebox{\zzzbar}
\sbox{\zzzbar}
{\setlength{\unitlength}{0.9em}
\begin{picture}(0.6,0.7)
\thinlines
\put(0,0){\line(1,0){0.6}}
\put(0,0.75){\line(1,0){0.575}}
\multiput(0,0)(0.0125,0.025){30}{\rule{0.3pt}{0.3pt}}
\multiput(0.2,0)(0.0125,0.025){30}{\rule{0.3pt}{0.3pt}}
\put(0,0.75){\line(0,-1){0.15}}
\put(0.015,0.75){\line(0,-1){0.1}}
\put(0.03,0.75){\line(0,-1){0.075}}
\put(0.045,0.75){\line(0,-1){0.05}}
\put(0.05,0.75){\line(0,-1){0.025}}
\put(0.6,0){\line(0,1){0.15}}
\put(0.585,0){\line(0,1){0.1}}
\put(0.57,0){\line(0,1){0.075}}
\put(0.555,0){\line(0,1){0.05}}
\put(0.55,0){\line(0,1){0.025}}
\end{picture}}

\newcommand{\bra}[1]{\langle{#1}|}
\newcommand{\ket}[1]{|{#1}\rangle}

\newcommand{\ena}{\end{eqnarray}}

\newcommand{\be}{\begin{equation}}
\newcommand{\ee}{\end{equation}}

\newcommand{\Tr}{{\rm Tr}}
\newcommand{\ktilde}{\tilde{k}}

\def\be{\begin{equation}}
\def\ee{\end{equation}}

\def\r{\rho}

\def\({\left (}
\def\){\right )}
\def\[{\left [}
\def\[{\right ]}

\def\tr{\mathrm{tr}}

\def\ba{\begin{eqnarray}}
\def\ea{\end{eqnarray}}

\def \r{{\bf r}}

\input amssym.def
\input amssym.tex

\newcommand{\bbibitem}[1]{\bibitem{#1}\marginpar{#1}}
\def\Bibitem#1{\bibitem{#1}%
  \smash{\hbox to0pt{\raise1ex\hbox{\tiny[#1]}\hss}}}

\def\Label#1{\label{#1}%
  \smash{\hbox to0pt{\raise1ex\hbox{\tiny[#1]}\hss}}}
\def\noLabels{\let\Label=\label}
\def\nobbibitem{\let\bbibitem=\bibitem}
 \def\noBibitem{\let\Bibitem=\bibitem}

\def\[{\left [}
\def\]{\right ]}
\def\({\left (}
\def\){\right )}

\def\r{\rho}

\def\r2{\sqrt{2}}

\def\bra{{\langle}}
\def\ket{{\rangle}}
\newcommand{\Xhat}{\hat{X}}

\def\Label#1{\label{#1}%
  \smash{\hbox to0pt{\raise1ex\hbox{\tiny[#1]}\hss}}}
\def\noLabels{\let\Label=\label}
\def\nobbibitem{\let\bbibitem=\bibitem}

\newcommand{\bea}{\begin{eqnarray}}
\newcommand{\eea}{\end{eqnarray}}

\newcommand{\beq} {\begin{equation}}
\newcommand{\eeq} {\end{equation}}
\newcommand{\bear}{\begin{eqnarray}}
\newcommand{\eear}{\end{eqnarray}}

\newcommand{\Ocal}{{\cal O}}

\newcommand{\beqn}{\begin{eqnarray}}
\newcommand{\eeqn}{\end{eqnarray}}

\def\bra{\langle}
\def\ket{\rangle}

\begin{document}

\begin{titlepage}
\begin{flushright}
arXiv:1807.07357\\
HIP-2018-25/TH
\end{flushright}
\vfill
\begin{center}
{\Large \bf %
Aspects of capacity of entanglement}

\vskip 10mm

{\large Jan~de Boer$^{a}$, Jarkko~J\"arvel\"a$^{b,c}$,  Esko~Keski-Vakkuri$^{b,c}$}

\vskip 7mm

$^a$ Institute for Theoretical Physics, University of Amsterdam, \\
\hspace*{0.15cm} Science Park 904, 1098 XH Amsterdam, The Netherlands\\
$^b$Department of Physics, P.O.Box 64, FIN-00014 University of Helsinki, Finland\\
$^c$Helsinki Institute of Physics, P.O.Box 64, FIN-00014 University of Helsinki, Finland\\

\vskip 3mm
\vskip 3mm
{\small\noindent  {\tt J.deBoer@uva.nl, jarkko.jarvela@helsinki.fi, esko.keski-vakkuri@helsinki.fi}}

\end{center}
\vfill

\begin{center}
{\bf ABSTRACT}
\vspace{3mm}
\end{center}

Many quantum information theoretic quantities are similar to and/or inspired by thermodynamic
quantities, with entanglement entropy being a well-known example. In this paper, we study a
less well-known example, capacity of entanglement, which is the quantum information theoretic
counterpart of heat capacity. It can be defined as the second cumulant of the entanglement spectrum and
can be loosely thought of as the variance in the entanglement entropy. We review the definition
of capacity of entanglement and its relation to various other quantities such as fidelity susceptibility
and Fisher information.

We then calculate the capacity of entanglement for
various quantum systems, conformal and non-conformal quantum field theories in various dimensions, and examine their holographic gravity duals.
 Resembling the relation between response coefficients and order parameter fluctuations 
in Landau-Ginzburg theories, the capacity of entanglement in field theory is related to integrated gravity fluctuations in the bulk.
We address the question of measurability,
in the context of proposals to measure entanglement and R\'enyi entropies by 
relating them to $U(1)$ charges fluctuating in and out of a subregion, for systems equivalent to non-interacting fermions.

From our analysis, we find universal features in conformal field theories, in particular the
area dependence of the capacity of entanglement appears to track that of the entanglement entropy.
 This relation is seen to be modified under
perturbations from conformal invariance. In quenched 1+1 dimensional CFTs, we compute the
rate of growth of the capacity of entanglement. 
The result may be used to refine the interpretation of entanglement spreading being carried by ballistic propagation of entangled quasiparticle pairs created
at the quench.

\end{titlepage}

\tableofcontents

\section{Introduction}

Entanglement entropy has been proven quite useful as a diagnostic of topological properties of ground states of quantum
many-body systems, for a recent review see {\em e.g.} \cite{Laflorencie:2015eck}. At the same time, starting with \cite{Ryu:2006bv}  
it has also been playing an instrumental role in 
understanding the connection between the geometry of space-time and strongly coupled field theories in the AdS/CFT 
correspondence, a recent extensive review is \cite{Rangamani:2016dms}. 

In this paper, we study another quantity associated to a reduced density matrix, namely the capacity of 
entanglement. It is defined in the same way as one defines heat capacity for thermal systems,
which was the original motivation to introduce this quantity in \cite{YaoQi} (see also \cite{Schliemann}).
Explicitly, if $\lambda_i$ are the eigenvalues of a reduced density matrix, entanglement entropy is defined
as $S_{EE}=-\sum\lambda_i \log\lambda_i$, and capacity of entanglement can be written as $C_E=\sum_i \lambda_i \log^2 \lambda_i -S^2_{EE}$. 
The latter
can also be thought of as the variance of the distribution of $-\log\lambda_i$ with probability $\lambda_i$, and
it is clear that it contains information about the width of the eigenvalue distribution of the reduced density matrix. 

Our main motivation was to 
understand the connection between capacity of entanglement and quantum fluctuations, and to {\emph e.g.} see whether
capacity of entanglement sheds any interesting light on the accuracy of the semi-classical approximation in gravity.
This connection turns out to be the following. As is well-known, the Ryu-Takanagi formula \cite{Ryu:2006bv} relates the entanglement entropy in a quantum field theory to the volume of a minimal surface ("RT surface") in a dual anti-de Sitter 
spacetime. If we include quantum gravity fluctuations of the RT surface, and integrate over them, the integral does indeed compute the capacity of entanglement in the quantum field theory \cite{Nakaguchi:2016zqi}. 
Thus, capacity of entanglement associates a ``width'' to the entanglement entropy: a measure of quantum gravitational effects. This relationship is a leading contribution to a more
complete sum of gravitational fluctuations. In \cite{Dong:2016fnf}, the RT surface was replaced by a domain wall with tension (a "cosmic brane"), including its complete 
gravitational backreaction. The volume of the brane in the backreacted geometry was shown to be equal to a variant of the R\'enyi entropies in the quantum field theory, coined ``modular entropy'' in \cite{Dong:2016hjy}. This result was used in  \cite{Nakaguchi:2016zqi}, independently the modular entropy was introduced 
and used to define the capacity of entanglement in the earlier work \cite{YaoQi,Schliemann}.

Entanglement entropy and capacity of entanglement are two features of the full entanglement spectrum \cite{LiHaldane}
(the set of eigenvalues of the reduced density matrix). There has been growing interest to
study the entanglement spectrum as a diagnostic of the phase structure of various systems. For our purposes, of special relevance are the studies of the entanglement spectrum for 1+1 dimensional
gapless \cite{CalLef, Lauchli:2013jga, Lundgren:2015iqm} and gapped \cite{CLR} systems.   
Originally it was also hoped that there would be a precise relation between the entanglement spectrum and the energy eigenvalue spectrum, but there are caveats \cite{Chandran}. Connections between the capacity of entanglement and the entanglement spectrum have been explored in \cite{Nakagawa:2017wis}.

Complete knowledge of all the R\'enyi entropies is in principle sufficient to 
recover the complete set of eigenvalues of the reduced density matrix,  the entanglement spectrum distribution, by a suitable integral transform. In this way,
\cite{CalLef} derived the spectrum for 1+1 conformal field theories in the ground state, with a finite length $\ell$ interval as the subsystem, and found
that the  spectrum is (under some assumptions) characterized by a universal function,  depending only on two parameters: the central charge $c$ of the theory and the largest
eigenvalue $\lambda_{max}$. Universality was also found in gapped systems \cite{CLR}, while the assumptions of \cite{CalLef} were further studied in \cite{Alba:2017bgn} and
the spectrum was interpreted as a reparametrization of the Cardy formula counting energy levels in a CFT.

While we will make a few remarks about the full entanglement spectrum, we will mostly restrict to the two lowest moments
or cumulants of the spectrum, which are precisely the entanglement entropy and capacity of entanglement. In terms of
the modular Hamiltonian $K_A = -\log \rho_A$ they are the expectation value and variance of $K_A$ respectively. As stated above, for systems of with a gravity dual, higher cumulants
capture the whole series of the quantum gravitational fluctuations about the RT surface, giving the complete entanglement spectrum or the R\'enyi / modular entropies.

Besides looking at gravitational fluctuations, we will in this work study the capacity of entanglement from many different points of view. We develop a more comprehensive list of its properties and
relations to other concepts of quantum information, not just
in the context of holographic gauge/gravity duality. We study a range of quantum systems from simple qubit systems and random pure states to spin chains, conformal and non-conformal quantum field theories,
with and without a gravity dual, and also non-equilibrium quenched systems. We are especially interested in identifying universal features. For example, we find evidence for an ``area
law'' for the capacity of entanglement in conformal theories. Curiously, in four-dimensional conformal field theories, 
in a fairly
natural regularization scheme, the ratio between the coefficients in front of the area term in entanglement entropy
and capacity of entanglement turns out to be precisely $a/c$, the ratio of the $a$ and $c$ anomaly coefficients,
for spherical entangling surfaces.
This area law suggests that most of the quantum fluctuations of the RT surface are located near the boundary of AdS and
this therefore does not seem to shed much light on the size of local bulk quantum fluctuations.

Another important observation is that systems where are all entanglement is carried by EPR pairs (pairs of qubits
that contribute $\log 2$ to the entanglement entropy) have zero capacity of entanglement. Therefore, whenever
we find that entanglement entropy and capacity of entanglement are approximately equal to each other, EPR pairs
are not a very good approximation of the quantum state. As we will show, randomly entangled pairs of qubits give a much
better description, and this can be used to e.g. sharpen the picture 
of ballistic propagation of entanglement carried by quasiparticles created at a quench.

The outline of this paper is as follows. Section \ref{Sec2} begins with some basic definitions and relations.  We begin with definitions of entropy, capacity of
entanglement, and modular entropy based on quantum information theory. We discuss
 the moments and cumulants of the entanglement spectrum / modular Hamiltonian, and show how to 
obtain them from the R\'enyi entropies. We then move to analogues of thermodynamic definitions of entropy and heat capacity applied to quantum entanglement,
and show agreement with the previous quantum information theoretic definitions. 
We move to review the definitions of  fidelity susceptibility and quantum Fisher information, and show how they are related to the capacity of
entanglement (related discussion is also in \cite{Banerjee:2017qti, Alishahiha:2017cuk}).
Finally, we map capacity of entanglement to the heat capacity
of a thermal CFT on a hyperbolic sphere, recovering the thermodynamic equality of  heat capacity and variance of entropy.

Section \ref{Sec3} discusses some properties of the capacity of entanglement.  We first derive an upper bound for it in quantum systems with a finite dimensional Hilbert space. 
We also recast this result as an estimate of the variance in classical information theory.
Addressing the measurability of the capacity of entanglement,  we follow the proposal to relate bipartite entanglement 
for non-interacting fermions with fluctuations of conserved $U(1)$ charges \cite{KRS,Klich:2008un,SRLH,SFRKLH,Song:2011gv}.
We show how the moments and cumulants of entanglement spectrum and $U(1)$ charges are related. We point out that an equality between 
the entanglement entropy and capacity of entanglement arises
naturally in this context in a large $N$ limit, at the leading level. We speculate how such a limit might arise in theories with a holographic bulk gravity dual, and consider an example of an eigenvalue
distribution giving rise to the equality $C_E = S_{EE}$.
We then interpret the capacity of entanglement holographically. We give a shorter derivation of
the bulk integral formula of  \cite{Nakaguchi:2016zqi}, and discuss its interpretation.

In Section \ref{Sec4}, we study how the capacity of entanglement depends on the initial quantum state of the complete system. In simple $n$-qubit examples, the entanglement entropy and the capacity of entanglement are in general not proportional.
For subsystems of random pure states, the approximate equality of the two can arise as ensemble averages, and we compute
the precise ensemble averages for several choices of the dimension of the parent Hilbert space where the pure states are 
defined and choices of dimension of the subsystem.   

In Section \ref{Sec5}, we study the capacity of entanglement in field theories. First we calculate it in 1+1 CFTs at equilibrium and in local and global quenches. In all cases,
the capacity of entanglement is equal to the entanglement entropy. For the non-equilibrium cases this implies that the
rate of growth of the capacity of entanglement follows that of $S_{EE}$. We comment on the interpretation
of ballistic spreading of entanglement. We then extend the computation to higher dimensions, following an alternative 
method based on stress-tensor correlators \cite{Perlmutter:2013gua,Hung:2014npa}.
After that, we study universality. We discuss some aspects of the entanglement spectrum and its (non)-universality.
Then we consider the possible dependence on regularization schemes. In generic quantum field theories and CFTs, 
the ratio $C_E/S_{EE}$ depends on the regularization scheme, but for CFTs with a holographic
gravity dual, the gravity interpretation provides a natural preferred regularization scheme.
With this scheme, for spherical entangling surfaces in $D=4$, the ratio is bounded as a 
consequence of the conformal collider bounds of \cite{Hofman:2008ar, Hofman:2016awc}. Restricting to theories
with holographic duals without higher derivative terms, the ratio turns out to be exactly equal to one.

Finally, in section \ref{Sec6}, we study how the capacity of entanglement behaves when conformal invariance is broken slightly. 
As a warm-up example, in the anisotropic Heisenberg XY spin chain, the equality of capacity and entropy 
emerges at critical domains. Moving away from criticality,
the capacity of entanglement and the entanglement entropy are no longer equal and
develop different subleading divergence structures. We then perform a more general analysis perturbing CFTs by relevant operators, following the strategy in
\cite{Rosenhaus:2014zza,Rosenhaus:2014woa,Rosenhaus:2014ula}.

We end with some concluding remarks.

\section{Definitions and relations}
\label{Sec2}

In this section, we introduce the concepts that will be the focus of later sections. As in thermodynamics, there will be two different routes to 
defining these concepts. 
A ``microscopic'' route is to follow the standard definitions of quantum information theory. The second, ``thermodynamic'' route, follows
from associating a Boltzmann distribution to the (reduced) density matrix and then formally applying thermodynamic definitions
to quantum entanglement. We will begin with the first route, then follow the second route, and finally establish
that the two different routes lead to the same concepts. Apart from some refinements, this section is mostly a review.

\subsection{Concepts of quantum information theory}\label{qidefinitions}

We start with a quantum system prepared into a pure or mixed state described by the density matrix $\rho$. Since we will mostly be interested in
bipartite entanglement between a subsystem $A$ and its complement, we assume that the density matrix is a reduced density matrix in the subsystem $A$.
However, this is just for terminological convenience, most of the concepts obviously apply more generally.

The most familiar concept quantifying the purity of the state  $\rho$, or the entanglement between the subsystem and its complement, is the entanglement
entropy 
\be
   S_{EE} = -\Tr (\rho \log \rho ) \ .
\ee
Introducing the modular Hamiltonian $K = -\log \rho$, the entanglement entropy becomes the expectation value of $K$,
\be
 S_{EE} = \Tr ( K\rho )= \bra K \ket \ .
\ee
Another much studied measure of entanglement is given by the R\'enyi entropies $S_\alpha$,
\be
S_\alpha = \frac{1}{1-\alpha}\log \Tr (\rho^\alpha) \ .
\ee
Recent work \cite{Dong:2016fnf} suggested that in addition to the R\'enyi entropies $S_\alpha$, it is interesting to study a modification, called 'modular entropy' in
\cite{Dong:2016hjy}, 
\beq
\label{Smodular}
  \tilde{S}_\alpha \equiv \alpha^2 \partial_\alpha \left( \frac{\alpha-1}{\alpha} S_\alpha \right) = -\alpha^2 \partial_\alpha \left( \frac{1}{\alpha} \ln \Tr \rho^\alpha \right)\ .
\eeq
The motivation in  \cite{Dong:2016fnf} comes from entanglement entropy in CFTs with a 
holographic gravitational dual, where the Ryu-Takayanagi formula relates the entanglement entropy to a volume of a dual minimal surface. In  \cite{Dong:2016fnf} 
it was found that the modular entropy also satisfies a holographic area law, with the interpretation\footnote{with  $\alpha$  being a natural number}
\beq
\label{Scosmicbrane}
 \tilde{S}_\alpha = \frac{{\rm Area(Cosmic\ Brane)}_\alpha}{4G_N} \ ,
\eeq
where the ``Cosmic Brane''  is a domain wall with tension $(\alpha -1)$ backreacting to the bulk geometry.  In the limit $\alpha \rightarrow 1$,
(\ref{Scosmicbrane}) reduces to the Ryu-Takayanagi formula.

In an earlier work \cite{YaoQi} (see also \cite{Schliemann}), (\ref{Smodular}) was used as a starting point to define a new quantity, capacity of entanglement $C_E$, to extend the thermodynamic relations found in quantum entanglement. In this paper we prefer to define it first in a microscopic,
quantum information theoretic way, as a property of the state $\rho$,
\be\label{ceqi}
 C_E (\rho) \equiv \Tr (\rho (-\log \rho)^2) - (-\Tr (\rho \log \rho))^2 \ .
\ee
From this definition, written in terms of the modular Hamiltonian $K$, the capacity of entanglement is equal to the variance (the second
cumulant) of $K$:
\be
C_{E} (\rho) = \bra K^2 \ket - \bra K \ket^2  \ .
\ee
The thermodynamic definition and its equality with (\ref{ceqi}) will be discussed in sections \ref{thermodef} and \ref{thermoqi}.

A complete set of data associated with $\rho$ is the entanglement spectrum \cite{LiHaldane},
the complete set of eigenvalues $\{ \lambda_m = e^{-\varepsilon_n}\}$ of the (reduced)
density matrix $\rho$ or the eigenvalues $\varepsilon_m$ of the modular Hamiltonian 
$K$, along with their multiplicities $g_m$. There are other ways to encode the data of the spectrum $\{g_m, \varepsilon_m\}$. One alternative
is to consider its  moments or cumulants. For that, one can define a generating function $k(\alpha )$ as the analytic continuation\footnote{In general, the analytic continuation may not be well defined for all $\alpha$, but we will need the continuation only into a small neighbourhood of $\alpha =1$.}
of the moments of the density matrix $\rho$,
\beq\label{spect}
   k(\alpha)  = \Tr \left(\rho^\alpha \right) %
= \sum_m g_m e^{-\alpha \varepsilon_m} \ . %
\eeq

By the usual rules, $k(\alpha)$ is a generating function for all the moments of the modular Hamiltonian:
\beq
 \bra K^n \ket = (-1)^n \left.\frac{d^nk(\alpha)}{d\alpha^n}\right|_{\alpha=1} =  \bra (-\ln \rho )^n \ket  = \sum_m g_m  e^{-\varepsilon_m}\varepsilon^n_m\ . 
\eeq
Formally, one may also refer to these as the moments of entanglement entropy\footnote{
In classical information theory, Shannon information $H=-\sum_i p(x_i)\log_2 p(x_i)$ quantifies the average information in messages composed of letters $x_i$ with probabilities $p(x_i)$, generated at a source. In this context, the higher moments of $H$ have a natural interpretation, characterizing the fluctuations in information. (A related concept is that of information loss in the evolution of chaotic dynamic systems, where the cumulants of $H$ \cite{Schlogl}, and their
generating function, the R\'enyi information \cite{GP} can be used to characterize dynamical chaos.) 
}, $S^n_{EE} = \bra K^n \ket $.
The partition function
is related to the (analytically continued)  R\'enyi entropies. Conversely, the analytic form of the R\'enyi entropies $S_\alpha$ determines $k(\alpha)=\exp [(1-\alpha)S_\alpha]$, from which the eigenvalue spectrum can be extracted  by
a suitable expansion or integral transformation (see {\em e.g.} \cite{Franchini:2007eu}).  

The logarithm of the generating function, 
\beq
\ktilde (\alpha) = \log k(\alpha) =(1-\alpha )S_\alpha \ ,
\eeq
is the generating function for the cumulants (connected correlators),
\beq\label{ktilde}
\bra K^n \ket_c = (-1)^n\left. \frac{d^n\ktilde (\alpha)}{d\alpha^n}\right|_{\alpha=1} \ .
\eeq
If the explicit form of the R\'enyi entropies $S_\alpha$ is known, through (\ref{ktilde}) by applying derivatives, we can then easily calculate
the capacity of entanglement $C_E$, which we defined in (\ref{ceqi}),
\begin{equation}
 C_E  = \bra K^2 \ket_c %
=\left.\frac{d^2[(1-\alpha)S_\alpha]}{d\alpha^2}\right|_{\alpha=1}  \ .\label{variance}
\end{equation}
The first derivative gives the first cumulant, which is $S_{EE}$.
To summarize, there are four alternative ways to represent the entanglement data: the entanglement spectrum, the R\'enyi entropies, 
the  moments $\bra K^n\ket$ or the cumulants $\bra K^n \ket_c$. It is possible to move from one description to the other.
For example, knowing all the moments allows to construct the generating function $k(\alpha)$, which determines the entanglement spectrum.
The different alternatives may be practical for highlighting different features of the entanglement data. The rest of the paper will focus on studying
what features are captured  by the second cumulant, the capacity of entanglement.

For pure states obviously all moments and cumulants vanish.
For maximally mixed states,
\beq
  \bra K^n \ket = (\ln d)^n \ ; \ \bra K^{n>1}\ket_c = 0 \ ,
\eeq
in a Hilbert space with dimension $d<\infty$. The %
 spectrum has only one $d$-fold degenerate eigenvalue $\varepsilon =\ln d$.

\subsection{Thermodynamical definitions}\label{thermodef}

In the previous section, we gave a quantum information theoretic definition and interpretation of the capacity of entanglement. In this section,
we follow the original definition, by following the analogues of thermodynamical relations applied to quantum entanglement. We begin
with the state $\rho = e^{-K}$ and introduce an inverse temperature $\beta$ and compute a partition function
\be
 Z_\rho (\beta) = \Tr (\rho^\beta) = \Tr (e^{-\beta K}) \ .
\ee
This is of course the same as the generating function $k(\alpha)$, with $\beta = \alpha$. The only reason for the two different notations is to make
a distinction between the thermodynamic and quantum information theoretic interpretations.
$Z(\beta)$ is a more natural notation in the thermodynamical context, while $k(\alpha)$ refers to the quantum information theoretic
context, hopefully this will not create confusion.

One can then proceed to introduce a free energy
\beq
F_\rho (\beta ) = -\frac{1}{\beta} \log Z_\rho (\beta) \ ,
\eeq
and an "internal" energy 
\begin{eqnarray}
  E_\rho (\rho) &=& \frac{\partial}{\partial \beta}(\beta F_\rho (\beta)) \nonumber \\
\mbox{} &=& \frac{\Tr (Ke^{-\beta K})}{\Tr (e^{-\beta K})} \equiv \bra K \ket_\beta
\end{eqnarray}
which, as seen in the last line, is just the ``thermal'' expectation value\footnote{Notation: $\bra (\cdot) \ket_\beta \equiv
\frac{\Tr ((\cdot )e^{-\beta K})}{\Tr (e^{-\beta K})}$.} of the modular Hamiltonian.

The thermodynamic definition of entropy satisfies the usual relation with the internal and free energy,
\begin{eqnarray}\label{Stherm}
S_\rho (\beta) &=& \beta^2 \frac{\partial F_\rho (\beta)}{\partial \beta} \nonumber \\
 \mbox{}  &=& \beta \bra K\ket_\beta + \log Z_\rho (\beta ) = \beta [E_\rho (\beta) - F_\rho (\beta)] \ .
\end{eqnarray}
Finally, one defines a heat capacity,
\beq
 C_\rho (\beta) = -\beta^2 \frac{\partial E_\rho (\beta)}{\partial \beta} = \beta^2 [\bra K^2 \ket_\beta - \bra K \ket^2_\beta ]\ .
\eeq

Except for the partition function, it is not manifestly clear how these thermodynamic definitions are related to the quantum information theoretic
definitions of section \ref{qidefinitions}. That is the topic of the next section.

\subsection{Relations between the thermodynamic and quantum information theoretic definitions}\label{thermoqi}

To begin with, from the density matrix $\rho = e^{-K}$ we construct a one-parameter family $\rho_\alpha$
of density matrices with proper normalization\footnote{In \cite{Nakaguchi:2016zqi}, they were
called escort density matrices, following the classical terminology of escort probability distributions.},
\beq\label{rhoalpha}
\rho_\alpha = \frac{e^{-\alpha K}}{\Tr(e^{-\alpha K})} = \left.\frac{e^{-\beta K}}{Z_\rho(\beta)}\right|_{\beta = \alpha} \ .
\eeq
The key idea here will be that when we apply the quantum information theoretic definitions of section \ref{qidefinitions} (which were given
for arbitrary states $\rho$) to the one-parameter family of states (\ref{rhoalpha}), we make contact with the thermodynamic definitions
with $\beta = \alpha$.

We consider first the entanglement
entropy. We observe that
\beq\label{SvN}
 S_{EE} (\rho_\alpha) = -\Tr  [\rho_\alpha \log \rho_\alpha]  = [\beta \bra K \ket_\beta + \log Z_\rho (\beta)]|_{\beta = \alpha}
 = S_\rho (\beta)|_{\beta=\alpha} \ ,
\eeq
verifying the relation between the two definitions of entropy.

On the other hand, we may consider the modular entropies $\tilde{S}_\alpha$ of the original state $\rho$. This gives another relation,
\begin{eqnarray}
\tilde{S}_\alpha (\rho) &=& -\alpha^2 \partial_\alpha \left(\frac{1}{\alpha} \log \Tr \rho^\alpha\right) \nonumber \\
   \mbox{} &=& [\beta \bra K \ket_\beta + \log Z_\rho (\beta)]|_{\beta = \alpha}
 = S_\rho (\beta)|_{\beta=\alpha} \ ,
\end{eqnarray}
establishing $\tilde{S}_\alpha (\rho)  = S_{EE} (\rho_\alpha) = S_\rho (\beta)|_{\beta=\alpha}$. 

Finally, we show that the quantum information theoretic definition of the capacity of entanglement $C_E$ (which was a property of an
arbitrary state $\rho$, characterizing the variance of its spectrum), applied to the one-parameter family $\rho_\alpha$, matches with
the thermodynamic definition:
\begin{eqnarray}
C_E(\rho_\alpha) &=& \Tr [\rho_\alpha (-\log \rho_\alpha)^2] -[-\Tr (\rho_\alpha\log \rho_\alpha )]^2 \nonumber \\
 \mbox{} &=& \left\{\beta^2 [ \bra K^2 \ket_\beta -\bra K\ket^2_\beta]\right\}_{\beta =\alpha} = C_\rho (\beta)_{\beta = \alpha} \ .
\end{eqnarray} 
This also establishes (setting $\alpha=1$) the quantum information theoretic counterpart of the thermodynamic response-fluctuation relation between heat capacity and variance of entropy: for a state $\rho$ the
entanglement heat capacity and the variance of entanglement entropy satisfy
\beq
  C_E(\rho) = \bra K^2 \ket_c = \Delta S^2_{EE} (\rho ) \ ,
\eeq
which we used as the ``microscopic'' definition of $C_E$  in (\ref{ceqi}).

Finally, later in the paper we will consider grand canonical ensembles of either identical bosonic or fermionic particles,
and there it will be convenient to use the expression of the capacity or variance in terms of the expectation values of 
the occupation numbers $n_l$ as a single sum,
\beq
C_E = \Delta S_{EE}^2 = \sum\limits_{l}(\log(1\pm n_l)-\log(n_l))^2(1\pm n_l)n_l \ , \label{variation-nl}
\eeq
where we choose the $+$ signs for bosonic and $-$ signs for fermionic systems.

\subsection{Variance, fidelity susceptibility and quantum Fisher information}\label{quinfo}

The capacity or variance %
(\ref{ceqi}), (\ref{variance}),
can also be related to other concepts of quantum information: fidelity susceptibility (FS) and quantum Fisher information (QFI). 
These concepts are important in quantum metrology, and can be used to in various contexts, such as Bose-Einstein condensation, temperature estimates, and quantum phase
transitions, see {\em e.g.} \cite{gu}, \cite{depasqualeetal}, and the recent extensive review \cite{braunetal} (and references therein). Here we give only a brief review for the purpose of making a comparison to the capacity of entanglement $C_E$.

Consider
deformations of a density matrix,  parametrized by a continuous parameter $\theta$. This essentially gives a one-parameter flow $\rho (\theta )$  in the state space. One measure of a ``distance'' between two density matrices $\rho, \sigma$ is the (Uhlmann) fidelity $F(\rho ,\sigma)$ \cite{uhlmann},
\beq
 F(\rho ,\sigma) = \Tr \sqrt{\sqrt{\rho}\sigma \sqrt{\rho}} \ .
\eeq
Then $F(\rho (\theta), \rho (\theta+\epsilon))$ can be used as a measure of how
much the density matrix changes along an infinitesimal flow $\theta+\epsilon$. Since the fidelity is at maximum value 1 at $\epsilon =0$, its Taylor series expansion is
\beq
  F(\rho (\theta), \rho (\theta +\epsilon)) = 1 - \frac{1}{2} \chi_F (\theta) \epsilon^2 + \cdots
\eeq
where the coefficient 
\beq
 \chi_F (\theta)  \equiv -\left.\frac{\partial^2 F(\rho (\theta),\rho (\theta +\epsilon))}{\partial \epsilon^2}\right|_{\epsilon=0}
\eeq
is called the fidelity susceptibility. For example, consider $\theta$ as a control parameter driving the ground state of a system across a quantum phase transition. Then the fidelity
susceptibility is an interesting observable, manifesting singular behavior that can be used to characterize the quantum phase transition . 

A slightly more general approach is to compare $\rho$ and $\rho + \epsilon \delta \rho$, and work in a basis where $\rho$ is diagonal with eigenvalues $\lambda_i$. Note however that in this
basis the perturbation $\delta \rho$ need not be diagonal. A brute force computation gives then
\beq
F(\rho, \rho +\delta \rho) = 1-\frac{\epsilon^2}{4} \sum_{ik}  (\delta\rho)_{ik} (\delta\rho)_{ki} \frac{1}{\lambda_i+\lambda_k} + {\cal O}(\epsilon^3)
\eeq
from which we read the fidelity susceptibility 
\beq \label{fidel}
\chi=\frac{1}{2} \sum_{ik}  (\delta\rho)_{ik} (\delta\rho)_{ki} \frac{1}{\lambda_i+\lambda_k} .
\eeq

We will shorty connect this to capacity of entanglement but before doing that we first discuss
another related concept, quantum Fisher information.  Let us specify a point $\theta =0$ and consider small deformations about it. In quantum measurements, one is often interested
in detecting the parameter $\theta$ by finding an observable $\hat{\theta}$, a locally unbiased estimator with the 
normalization
\beq
 \frac{\partial}{\partial \theta} \Tr (\rho (\theta) \hat{\theta})|_{\theta =0} = 1 \ . 
\eeq
The accuracy of the estimates
 of $\theta$ is characterized by the variance $\Delta \theta^2$, which is bounded from below by the quantum version of
the Cram\'er-Rao bound \cite{qcramerrao},
\beq
 \Delta \theta^2 \geq \frac{1}{N g(\theta)}
\eeq
where $N$ is the number of samples and the lower bound is characterized by the quantum Fisher information $g(\theta)$. 
To define it, consider first the so-called symmetric logarithmic derivative $L(\theta)$ of $\rho (\theta)$, defined by
\beq
\frac{\partial\rho (\theta)}{\partial\theta} = \frac{1}{2} (L\rho+\rho L) \ .
\eeq
The quantum Fisher information $g(\theta )$ is then defined as
\beq
g(\theta) = {\rm Tr} (\rho (\theta) L^2(\theta)) = {\rm Tr} \left( \frac{\partial\rho (\theta)}{\partial\theta} L(\theta )\right) . 
\eeq
 
For an explicit formula, consider again a small deformation $\rho +\epsilon \delta \rho$, in a basis where $\rho$ is diagonal (but $\delta \rho$ need not be). 
Then solve the equation
\beq
\delta\rho = \frac{1}{2} (L\rho+\rho L)
\eeq
for $L$,  in terms of the eigenvalues $\lambda_i$ of $\rho$ we get 
\beq
L_{ij} = \frac{2}{\lambda_i+\lambda_j} (\delta\rho)_{ij}  \ .
\eeq
It is now straightforward to establish an equality relating the Fisher information to the fidelity susceptibility (\ref{fidel}), 
\beq\label{QFI}
g= {\rm Tr}(\rho L^2) = 4\chi \ .
\eeq

Coming back to the variance $\langle K^2 \rangle_c$, let us consider a particular deformation of the density matrix $\rho = e^{-K}$ by deforming in the direction of (imaginary) modular flow,
\beq\label{sflow}
   \rho (\theta) = e^{-(1+\theta) K} / (\Tr~e^{-(1+\theta )K}) \ ,
\eeq
in other words the family of escort matrices (\ref{rhoalpha}). Computing the fidelity $F(\rho (0),\rho (\theta))$ to second order in $\theta$ (the normalization factor
must also be expanded), and setting $\theta=0$, yields
\beq\label{relations}
  C_E=\langle K^2 \rangle_c =  4 \chi_F (0) = g  \ ,
\eeq
{\em i.e.}, the capacity $C_E$ is equal to the fidelity susceptibility for this particular flow and the quantum Fisher information for estimating the parameter $\theta$. 
Alternatively, expanding $\rho(\theta) \approx \rho(0)+\theta \delta \rho + {\cal O}(\theta^2)$, if $\rho (0)$ is diagonal with eigenvalues $\lambda_i$, 
then $\delta \rho$ is diagonal
with eigenvalues $\lambda_i\log \lambda_i - \lambda_i \sum_j \lambda_j \log \lambda_j$, and then (\ref{fidel}), (\ref{QFI}) yield (\ref{relations}).

The equality (\ref{relations}) is well-known, see {\em e.g.} \cite{Alishahiha:2017cuk}. 
In a finite temperature system, when the density matrix is that of a canonical ensemble, the above quantities are called
the thermal fidelity susceptibility {\em etc.}, and also equal to the heat capacity \cite{thermofid}.
However, we emphasize that the first equality in (\ref{relations}) assumes the particular flow (\ref{sflow}). For example, suppose that
we deform the modular Hamiltonian by a perturbation $V$, to consider a flow
\beq
    \tilde{\rho}(\theta) = e^{-K-\theta V}/(\Tr~e^{-K-\theta V}) \ .
\eeq
This leads to a more complicated expansion of the fidelity, since in general $V,K$ do not commute, and the second equality in (\ref{relations}) does not hold.

\subsection{Relation to thermal heat capacity}

So far we have mostly been keeping the discussion at a general level, without specifying the origin of the quantum state $\rho$. In the remainder of the paper, we will
narrow our focus to bipartite entanglement in quantum systems, where the system is first prepared to a generic quantum state $\rho_0 $, then partitioned into a subsystem
$A$ entangled with its complement $B$, and then study the reduced density matrix $\rho_A=\Tr_B \rho_0$. The entanglement entropy is the von Neumann entropy
of $\rho_A$. In this section, 
we consider specifically the entanglement entropy for CFT initial vacuum states and spherical entangling surfaces, in the context of \cite{Casini:2011kv, Hung:2011nu}. 
In these papers, the authors discuss the interpretation of EE and R\'enyi entropies, by mapping the causal development of the enclosed ball $A$ to a hyperbolic
cylinder $\mathbb{R} \times H^{d-1}$, where the CFT in the vacuum in the original spacetime becomes a CFT in $H^{d-1}$ in a thermal bath, with temperature
\beq\label{T0}
   T_0 = \frac{1}{2\pi R}\, ,
\eeq
where $R$ is the radius of the original entangling sphere and the curvature scale of $H^{d-1}$.
The entanglement entropy is then mapped to standard thermal entropy $S_{therm}(T_0)$ of the CFT in the thermal bath, similarly the R\'enyi entropies are mapped
to those of the thermal ensemble. 

We consider the capacity of entanglement $C_E$ %
of the ball-like subsystem.
We show that it maps to the standard thermal heat capacity $C_{therm}$ of the thermal CFT. We start from\footnote{We use the notation $\alpha$ where \cite{Hung:2011nu} uses $q$.}  (\ref{variance}), but do not yet take the limit
$\alpha \rightarrow 1$ but consider
\beq
 C(\alpha ) = \alpha^2 \partial^2_\alpha \ln \Tr \rho^\alpha_A =  \alpha^2 \partial^2_\alpha [(1-\alpha)S_\alpha]\, ,
\eeq
where $S_\alpha$ is the R\'enyi entropy. On the other hand, it is related to the free energy $F$ on the hyperbolic sphere  by  \cite{Hung:2011nu}
\beq
 S_\alpha = \frac{\alpha}{1-\alpha} \frac{1}{T_0} [F(T_0)-F(T_0/\alpha)]  \ ,
 \eeq
so
\beq
 C(\alpha ) = \alpha^2 \partial^2_\alpha\left[-\frac{\alpha}{T_0}F\left(\frac{T_0}{\alpha}\right)\right] \ .
\eeq
Writing $T_\alpha \equiv T_0/\alpha$ and $\beta_\alpha = 1/ T_\alpha$, we get
\beq
C(\alpha ) = %
-\beta^2_\alpha \frac{\partial }{\partial \beta_\alpha} E(T_\alpha) 
= \frac{\partial E(T_\alpha)}{\partial T_\alpha} \ .
\eeq
Thus,
\beq
C_E = \lim_{\alpha\rightarrow 1} C(\alpha ) = \frac{\partial E(T_0)}{\partial T_0} = C_{therm} \ ,
\eeq
{\em i.e.}, the capacity of entanglement $C_E$ of $\rho_A$ becomes the usual heat capacity $C_{therm}$ of the thermal CFT on the hyperbolic sphere, as was expected\footnote{A precursor of such a connection was discussed in \cite{Brustein:2003kc}, which introduced a heat capacity of entanglement across a planar entangling surface, relating it by a coordinate transformation to the thermal heat capacity of Rindler radiation, also finding a linear area scaling.}. The temperature is parameterized by the radius of the entangling sphere, (\ref{T0}).
This also guarantees that the holographic
dual interpretation with a topological black hole in the bulk will be the heat capacity (associated with the horizon) of the black hole. 
Then, dialing the temperature of the black hole translates back in the original spacetime to the imaginary modular flow among the escort matrices accompanying the reduced density matrix $\rho_A$ of the subsystem A. 
 
It should be noted that in Appendix A, \cite{Hung:2011nu}
considered various inequalities satisfied by the R\'enyi entropy, and interpreted one of them (their (A.4)) as a result of proportionality to a heat capacity which is positive. These inequalities
were studied and proven in \cite{Nakaguchi:2016zqi}.

\paragraph{Interim summary.} We have established a string of equalities between the variance of the entanglement spectrum $\bra K^2\ket_c$, the variance of entanglement entropy,
the capacity of entanglement, a reduced fidelity susceptibility and quantum Fisher information,
\beq\label{equalities}
\bra K^2\ket_c = \Delta S^2_{EE} = C_E \stackrel{*}{=} 4\chi_F = g \ .
\eeq 
However, the third equality (marked with an asterisk) holds for a particular (modular) flow among density matrices. 
Hereafter, we will mostly adopt the notation $C_E$, and we will be interested in
comparisons to the entanglement entropy $S_{EE}$, in a given state $\rho$.

\section{Properties of capacity of entanglement}
\label{Sec3}

We have introduced the cumulants of the modular Hamiltonian, and interpreted the second cumulant as the capacity of entanglement or the variance of entanglement entropy.
In this section we study some properties of these concepts.

\subsection{An upper bound on the capacity of entanglement}

In a system with a finite Hilbert space of dimension $N$, the von Neumann entropy takes values between 0 and $S_{max}= \ln N$. 
It is interesting to derive similar bounds for the higher cumulants of the spectrum of $\rho$. In this section we derive an upper bound for the capacity of entanglement $C_E=\bra K^2\ket_c$.

We may assume that we have diagonalized the density matrix. We denote the eigenvalues $\lambda_i$, with the degenerate eigenvalues also carrying separate labels. 
There are two types of questions to consider: (i) for a given entropy $S$, what is the maximum variance $\bra K^2\ket_c$, or (ii) find the maximum of $\bra K^2\ket_c$ with no restriction on
the entropy. 

We start with the problem (i). It leads to the variational problem to extremize the functional
\beq
F = \sum_i \lambda_i (\log\lambda_i)^2 + A(S-\sum_i \lambda_i\log\lambda_i) + \sum B_i \theta(-\lambda_i) + C(\sum_i\lambda_i-1)\, ,
\eeq
where the Lagrange multipliers $A$, $B_i$ and $C$ enforce that the entropy takes a given value, 
that the $\lambda_i$ are non-negative, and that the sum of the eigenvalues is equal to one, respectively.
Note that $F$ is not equal to $\bra K^2\ket_c$, instead $F= \bra K^2\ket_c + S^2$, but since $S$ is fixed extremizing $F$ is the
same as extremizing $\bra K^2\ket_c$.

Differentiating $F$ with respect to  $\lambda_i$ yields
\beq
\log^2\lambda_i + (2-A)\log\lambda_i -A  - B_i\delta(\lambda_i) +C =0 .
\eeq
We see that if $\lambda_i \neq 0$, this is a quadratic equation for $\log\lambda_i$ with at most
two solutions for $\lambda_i$. In addition, some of the $\lambda_i$ could be zero, for which this
equation is somewhat ill-defined, but can formally be solved with $B_i=0$ and $C$ infinite.

We now first move on to the alternative problem (ii), which turns out to lead to a simple result, before turning back
to (i). If we do not fix $S$, we need to maximize
\beq
\tilde{F}=\sum_i \lambda_i (\log\lambda_i)^2 -(\sum_i \lambda_i\log\lambda_i)^2 + \sum B_i \theta(-\lambda_i) + C(\sum_i\lambda_i-1) .
\eeq
Differentiating with respect to $\lambda_i$ now leads to
\beq
\log^2\lambda_i + 2\log\lambda_i  - 2(1+\log\lambda_i) (\sum_i \lambda_i\log\lambda_i)
+ B_i\delta(\lambda_i) +C =0 .
\eeq
Also in this case, having three different non-zero values among the $\lambda_i$ would lead to a contradiction,
as all three would be solutions of a quadratic equation for $\log\lambda_i$ with the same coefficients.

We therefore conclude that for either optimization problem, the non-zero eigenvalues can take at most two different values.
Since a zero eigenvalue does not contribute to either entanglement entropy or the variance, we will ignore
the zero eigenvalues for now. 

We will denote the two non-zero eigenvalues by $\lambda_1$ and $\lambda_2$. 
If we call the corresponding occupation numbers $n_1$ and $n_2$, then $n_1\lambda_1 + n_2 \lambda_2=1$. We can
express $n_1$ and $n_2$ in terms of $\lambda_1$ and $\lambda_2$ and when we do this we find 
\beq
C_E = \bra K^2\ket_c =-(S+\log\lambda_1)(S+\log\lambda_2)\, ,
\eeq
which shows that in order to maximize $C_E$, we should make $\lambda_1$ and $\lambda_2$ as different as possible. 

It will be convenient to introduce the variables 
\beq 
x=n_1\lambda_1 \ , \ y = \frac{n_1}{n_2}
\eeq
so that we can express all quantities in terms of $x,y$ and $N=n_1+n_2$. We find that 
\beq
S =\log N - x \log \frac{x}{y(1-x)} - \log(1-x)(1+y)
\eeq
and 
\beq
 C_E= \bra K^2\ket_c  = x(1-x) \left[ \log \left( y\frac{1-x}{x}\right) \right]^2 \ .
\eeq

Let us try to maximize $\bra K^2\ket_c$.
We notice that first of all we want to minimize $y$. The minimum value would be $y=0$, but this is just the maximally entangled state,
for which the capacity of entanglement vanishes. Instead, we set $y=1/(N-1)$. Then, the maximum of $\bra K^2\ket_c$ is found at $x_N$, which is the solution of
\beq\label{transc}
  \log \left( \frac{1-x}{x}\right) -\frac{2}{1-2x} - \log (N-1) =0 \ .
\eeq
We then find an upper bound for the variance,
\beq
\bra K^2\ket_c \leq \frac{4x_N(1-x_N)}{(1-2x_N)^2} \ .
\eeq
Solving the transcendental equation (\ref{transc}) iteratively, we obtain for large $N$
\beq
 x_N \approx \frac{1}{2}+\frac{1}{\log (N-1)} + \cdots 
\eeq
and
\beq
  C_{E, \rm max}=\bra K^2\ket_{{c, \rm max}} \approx \frac{1}{4}\log^2 N +1+\cdots  \approx \frac{1}{4} S_{{\rm max}}^2 \ .
\eeq
Notice that we did not keep $S$ fixed when we varied $x$ and $y$. However, for the above
values of $x$ and $y$ the entropy for large $N$ remains of the order $S  \sim\log N$
so we expect that it remains true that $C_{E,\rm max}$ is proportional to $S^2_{\rm max}$
(with a different numerical constant) if we were to vary $\bra K^2\ket_c$ keeping $S$ fixed
and of the order $\log N$. It would be interesting to know if states with maximum capacity of entanglement have
special properties, but we have no specific insights at this point. \\

\noindent
Formally, the inequality 
\beq\label{ineq}
 C_E = \bra K^2 \ket_c = \Delta S^2_{EE} \leq \frac{1}{4} (S_{\rm max})^2
\eeq
resembles Popoviciu's inequality on variances \cite{Popoviciu}, which states that the variance of a
random variable $X$ with a range from $\inf(X)=m$ to $\sup(X)=M$ satisfies 
\beq\label{pop}
\Delta X^2  \equiv {\rm var}(X) \leq \frac{1}{4} (M-m)^2 \ .
\eeq
If we were to interpret the entropy $S_{EE}$ as a random variable with
the upper bound $S_{\rm max}$ and the lower bound 0,  the inequality (\ref{ineq}) would indeed superficially look like a special case of Popoviciu's inequality. However, the spectrum
of the modular Hamiltonian depends on the probability distribution, and has no finite upper bound. To clarify, we recast our result in the language of classical information theory.
 Let $X$ be a random variable with
$N$ outcomes $\{x_1,\ldots ,x_N\}$, with a discrete probability distribution $p(x_i)$. Then move to a new random variable, the information content $I(X) = -\log_2 p(X)$. The expectation
value is the Shannon information, $H=E(I(X))$, with $0\leq H\leq H_{\rm max} = \log_2 N$. The same variational calculation as we performed above, changing the notation, gives an estimate for the variance of $I(X)$
\beq\label{infoineq}
\Delta I(X)^2 \equiv {\rm var}(I(X))  \leq \frac{1}{4} (H_{\rm max})^2 \ .
\eeq
This is not a special case of Popoviciu's inequality: unlike for $X$, for the random variable $I(X)$ the supremum depends on the probability distribution $p(x)$, with $\sup I(X) = \infty$.
Instead, what appears in (\ref{infoineq}) is the supremum of the expectation value. We are unaware of whether (\ref{infoineq}) is a new result in classical information theory, or previously known.

Note also that the upper bound applies for a finite system. In the thermodynamic limit, with the system size 
becoming infinite, it would be interesting to study if the capacity of entanglement displays interesting non-analyticity {\em e.g.} when the system 
is undergoing a quantum phase transition.

\subsection{Bipartite particle fluctuations and entanglement fluctuations}
\label{Nfluct}

In this section, we discuss scenarios where it is natural to consider fluctuations in entanglement entropy and its cumulants. We consider entanglement in a bipartitioned system.  
There has been growing interest in developing strategies to measure entanglement. An
interesting proposal relates entanglement in a bipartitioned system (between the subsystem $A$ and its complement) to fluctuations of a conserved $U(1)$ charge, such as the particle number,
across the partition boundary \cite{KRS,Klich:2008un,SRLH,SFRKLH,Song:2011gv}, for systems that can be mapped
to non-interacting fermions. The particle number in the 
subsystem $A$, $N_A$, thus becomes
a random variable, and so does the entropy as entangled charges wander in and out. There are detailed results relating the fluctuations 
in the particle number $N_A$ in the subregion $A$ to the entanglement entropy $S_{EE}$ and even to the R\'enyi entropies $S_\alpha$. The probability distribution of the 
random number $N_A$ is characterized by its cumulants $n_m$, defined by 
\beq
 n_m = (-i\partial_\lambda)^m \log \chi (\lambda)|_{\lambda=0}\, ,
\eeq
with the generating function
\beq
\chi (\lambda) = \bra e^{i\lambda N_A} \ket\, ,
\eeq
where the expectation value is computed with respect to the initial state of the full system. Thus, for example the second cumulant $n_2$ (also called the fluctuations) is
\beq
 n_2 = \bra N^2_A \ket - \bra N_A\ket^2 \ .
\eeq

Given the relation between entanglement and particle number fluctuations, we expect the probability distributions of these two random variables to be related, and in particular to
be able to express all the cumulants of the entanglement entropy (the modular Hamiltonian) in terms of the cumulants of the particle number. Let us show this in more detail.  We emphasize that this only applies to 
systems that can be mapped to non-interacting fermions. In this context, the entanglement R\'enyi entropies have the following series expansion in terms of the $n_m$:
\beq\label{sC}
  S_{\alpha}(A) = \sum^\infty_{k=1} s^{(\alpha)}_k n_{2k}
\eeq
with the coefficients
\beq
  s^{(\alpha)}_k = \frac{(-1)^k(2\pi )^{2k}2\zeta (-2k,(1+\alpha)/2)}{(\alpha -1)\alpha^{2k}(2k)!}\, ,
\eeq
where $\zeta(s,a)$ is the Hurwitz zeta function. Note that only the even cumulants $n_{2k}$ appear in the series. In principle, the series is valid for real values of $\alpha >1$, but
one has to be careful about its convergence. In \cite{CMV}, it was shown that, for systems equivalent to noninteracting fermion gases, the series (\ref{sC}) is well behaved, because
the first cumulant $n_2$ contributes the leading $N^{(d-1)/d}\log N$ asymptotic behavior, while the  higher cumulants $n_{k>2}$ contribute only subleading behavior, $N$ being the total particle number.  Thus the series essentially truncates. 

In this case, we can view $\tilde{k}(\alpha )=(1-\alpha)S_{\alpha}(A)$ as a well behaved generating function for the cumulants of the entanglement entropy / modular Hamiltonian.
In particular, the entanglement entropy and the capacity of entanglement have the expansions
\begin{eqnarray}
 && S_{EE}(A) = \sum^\infty_{k=1} \frac{(2\pi)^{2k}|B_{2k}|}{(2k)!} n_{2k} = \frac{\pi^2}{3} n_{2} +\frac{\pi^4}{45} n_{4} + \frac{2\pi^6}{945}n_6 +\cdots\, ,  \\
 &&  C_E(A) = \frac{\pi^2}{3} n_2+ \sum^{\infty}_{k=2} \frac{2(2\pi)^{2k}|B_{2k}|}{(2k-1)!} n_{2k} = \frac{\pi^2}{3} n_2 +\frac{8\pi^4}{45} n_4 +\frac{24\pi^6}{945} n_6+\cdots \ , \nonumber
\end{eqnarray}
in the above $B_{2k}$ are Bernoulli numbers. From the results, we immediately see that if the particle fluctuations are Gaussian distributed ($n_{m>3}=0$), the entanglement
entropy is equal to its variance. Furthermore, in \cite{CMV} it was noted that in the $N\rightarrow \infty$ limit the R\'enyi entropies remarkably satisfy
\beq
   \frac{S_\alpha (A)}{n_2} = \frac{\pi^2}{6} (1+\alpha^{-1}) + \cdots\, ,
\eeq
where $+\cdots$ indicates corrections that vanish. A consequence of this is that 
\beq
C_E(A)=S_{EE}(A)
\eeq
 in the $N\rightarrow \infty$ limit. Note that the dimension $d$ of the system  was generic
in the derivation. However, here the size of the subregion $A$ is independent of $N$ and is kept fixed. 

As a diversion, let us make some very speculative but perhaps inspirational remarks regarding  
systems that have holographic gravity duals.  In this setting, we might make the following argument for invoking a large $N$ limit and Gaussian statistics. 
Holographic gauge/gravity duality is simplified in
a limit where the bulk gravity system is classical, corresponding to a large ${\cal N}$ limit in the dual theory at the AdS boundary, where ${\cal N}$ is the number of constituent degrees of freedom.
In the dual theory, one could then focus on the average number of degrees of freedom fluctuating in the subregion $A$,
\beq
    N_A  = \frac{1}{{\cal N}} \sum^{{\cal N}}_{i=1} N^{(i)}_A \ , 
\eeq
where $N^{(i)}_A$ is the particle number of the species $i$ in the subregion $A$. Viewing each $N^{(i)}_A$ as an independent random variable, not necessarily having identical 
distributions, under suitable conditions ({\em e.g.} Lyapunov conditions), the central limit theorem applies and in the large ${\cal N}$ limit the average particle number $N_A$ becomes
Gaussian distributed. If the quantum entanglement is attributed to the average particle number fluctuations about the subregion, from above one could deduce that $C_E=S_{EE}$. An additional complication is that in gauge/gravity duality, one typically works in the large 't Hooft coupling limit so that the bulk spacetime is weakly curved. The
dual theory on the boundary is then strongly coupled, so the concept of "particles" is lost and the definition of a "particle number" becomes less clear. Another caveat is that the definition
of entanglement for gauge invariant observables is more complicated \cite{Ghosh:2015iwa,Soni:2015yga}.

In Section \ref{XYsec}, we study a system which can be mapped to noninteracting fermions. 
We consider ground state entanglement in a Heisenberg XY spin chain at different phases, with the ground state adjusting to the changes in the parameters. 
Its R\'enyi entropies were computed  in \cite{Franchini:2007eu,Franchini:2010kq}, although they worked in a double
scaling limit $N\rightarrow \infty$, $L\rightarrow \infty$ with $L/N$ fixed, where $L$ is the size of the subsystem. Therefore, the above large $N$ result from particle number fluctuations
is not directly applicable. In particular, we will find that the entanglement entropy is not always equal to the capacity
of entanglement. 
This system will  also work as an introduction to later sections where we study CFTs and their perturbations with relevant
operators.

\subsection{An simple example where capacity equals entropy}

The large $N$ limit in the previous section led to the equality $C_E =S_{EE}$. Later, we will arrive several times
at the same result when we discuss entanglement in CFTs.  
It is interesting to ask what probability distributions could lead to such an equality. Here is one concrete example. 
It will be convenient to write $\lambda=e^{-E}$ and to use a density of states $\rho(E)$. 
For a finite dimensional density matrix with eigenvalues $\lambda_i$, the density of states would be
$\rho(E)=\sum_i \delta(E+\log\lambda_i)$. 
Then,
\begin{eqnarray}
\int dE \rho(E) e^{-E} & = & 1\ , \nonumber \\
\int dE E \rho(E) e^{-E} & = & S_{EE}\ , \nonumber \\
\int dE E^2 \rho(E) e^{-E} & = & C_E - S_{EE}^2\, .
\end{eqnarray}
If we take, for example, $\rho(E)=E^k/k!$, we obtain $C_E=S_{EE}=k+1$, which is indeed
an example where $C_E=S_{EE}$.

\subsection{Gravity dual of the capacity of entanglement}

Results in \cite{Dong:2016fnf} can be used to find the gravity dual of
the capacity of entanglement. This was done
in \cite{Nakaguchi:2016zqi},  but here we present a simplified derivation of the result, and point out
some additional properties. 

Starting from the modular entropy (\ref{Smodular}), we can rewrite the capacity of entanglement as a first derivative,
\beq
C_E = \Delta S^2_{EE}= -\left. \frac{d\tilde{S}_{\alpha}}{d\alpha} \right|_{\alpha=1} \ .
\eeq
The modular entropy $\tilde{S}_{\alpha}$ is given by the area of a suitably backreacted
cosmic brane as in (\ref{Scosmicbrane}) and to determine that
we need to consider the action for a brane coupled to gravity
\beq \label{aux11}
I=-\frac{1}{16 \pi G_N} \int \sqrt{g} R + I_{\rm matter} + \frac{\alpha-1}{4 \alpha G_N}\int_{\rm brane}\sqrt{h}
\eeq
with $h$ the induced metric on the brane.
To leading order, the cosmic brane is just a minimal surface. To first subleading order,
we need to take backreaction from the tension of the brane into account. The brane 
yields a source for the metric, and to first order in the source this affects the metric
everywhere in spacetime through the bulk graviton propagator. It could also affect the matter
fields to first order, in case the kinetic terms for the graviton and matter fields mix in this background.

At higher order, we would would
need to include more complicated Witten diagrams to understand the backreaction. 
If we only consider linear metric fluctuations, the location of the minimal surface does not change,
only its area changes. This is because the minimum area surface
is an extremum of the area functional and therefore its location does not change under
first order perturbations. 

The first-order variation of the area is given by
\beq \label{vara}
\delta \frac{{\rm Area}}{4G_N} = \frac{1}{4G_N} \int \sqrt{h} \frac{1}{2} h^{ij}\delta h_{ij}\, .
\eeq
The field equation of (\ref{aux11}) is
\beq \label{fe1}
-\frac{1}{16\pi G_N} (R_{\mu\nu} - \frac{1}{2} R g_{\mu\nu}) +\frac{1}{2} T^{\rm matter}_{\mu\nu}
-\frac{\alpha-1}{8 \alpha G_N} h_{ij} \delta({\rm brane}) =0 \ ,
\eeq
where a simplified notation has been used. A more precise
way to write the last term, using the explicit expression for the induced metric $h_{ij}$ in terms of the
embedding $x^{\mu}(\sigma^i)$ of the surface,
\beq
h_{ij} = g_{\mu\nu}(x^{\mu}(\sigma^i)) \frac{\partial x^{\mu}}{\partial \sigma^i}
\frac{\partial x^{\nu}}{\partial \sigma^j} \ ,
\eeq
reads
\beq
-\frac{\alpha-1}{8 \alpha G_N} \int d^{d-1}\sigma_i \frac{\sqrt{h}}{\sqrt{g}}
h^{ij} g_{\mu\rho} \frac{\partial x^{\rho}}{\partial \sigma^i}
\frac{\partial x^{\eta}}{\partial \sigma^j} g_{\eta\nu} \delta(x^{\mu}- x^{\mu}(\sigma^i)).
\eeq
In order to not clutter the notation, we will keep using the imprecise notation in (\ref{fe1})
but write the final answer in a more accurate way.

We need to solve (\ref{fe1}) to first order in $(\alpha-1)$. To leading order, the metric obeys the field
equations in the bulk, and we need to linearize around the background in order to find the first
order variation, $g_{\mu\nu}=g^{\rm background}_{\mu\nu}  + \delta g_{\mu\nu}$
\beq \label{fe2}
-\frac{1}{16\pi G_N} {\cal D}_{\rm background}\delta g_{\mu\nu}  -\frac{\alpha-1}{8 n G_N} g^{\rm background}_{ij} \delta({\rm brane})
+{\cal O}((\alpha-1)^2) =0
\eeq
with ${\cal D}$ some second order operator representing the kinetic term of the graviton.
Therefore, if $G$ is the graviton propagator in the background, we get
\beq
\delta g_{\mu\nu}(x) = -\frac{1}{8 G_N}\int dx' \sqrt{g(x')} G_{\mu\nu}^{\alpha\beta}(x,x')\frac{\alpha-1}{\alpha} g^{\rm background}_{\alpha\beta}(x')
 \delta({\rm brane})\, .
\eeq
Inserting this in the variation (\ref{vara}), we are left with a double integral over the 
minimal surface
\beq
\label{dsfin}
C_{E} = \frac{1}{64 G_N^2} \int dx \sqrt{g(x)} \int dx' \sqrt{g(x')} 
g^{ij} G_{ij}^{kl}(x,x') g_{kl}\, .
\eeq
Notice that the indices of the graviton propagator are contracted with the metric on the brane, in
other words we are effectively taking a trace, but only in the directions along the brane. 

A more precise version of (\ref{dsfin}) in terms of the embedding $x^{\mu}(\sigma^i)$ is
\beq
\label{dsfin2}
C_{E} = \frac{1}{64 G_N^2} \int d^{d-1}\sigma \sqrt{h(\sigma)} \int d^{d-1}\sigma' \sqrt{h(\sigma')} 
h^{ij} G_{ij;kl}(x(\sigma),x(\sigma')) h^{kl}  
\eeq
with
\beq
G_{ij;kl}(x(\sigma),x(\sigma')) = G_{\mu\nu;\rho\eta}(x(\sigma),x(\sigma')) 
\frac{\partial x^{\mu}}{\partial \sigma^i}
\frac{\partial x^{\nu}}{\partial \sigma^j}
\frac{\partial x^{\rho}}{\partial \sigma'^k}
\frac{\partial x^{\eta}}{\partial \sigma'^l}\, .
\eeq
This result was derived in \cite{Nakaguchi:2016zqi}, via a different route. 

We note in passing that the result (\ref{dsfin2}) looks very familiar when phrased in terms of the (reduced) fidelity susceptibility associated with the flow (\ref{sflow}) as in Section 2. Then\footnote{absorbing the normalization factor into the graviton propagator},
\beq
   \chi_F =  \int_\Sigma d\vec{x} \int_{\Sigma} d\vec{x}'  g^{ij} g_{kl} G^{kl}_{ij}(\vec{x},\vec{x}') \ \ ,
\eeq
a relationship between response coefficient and fluctuations resembling that of the magnetic susceptibility and order parameter fluctuations in Landau-Ginzburg theory,
\beq
 \chi = \int~d\vec{x} \int~d\vec{x}'  \delta^{ij} C_{ij} (\vec{x},\vec{x}') \, ,
\eeq
with $C_{ij}(\vec{x},\vec{x}') ) = \langle m_i(\vec{x})m_j(\vec{x}')\rangle_c $. The analogue is satisfying, viewing  bulk gravity as an effective theory for the quantum theory
on the boundary.

One might have thought that one can simply decouple this piece of the graviton propagator by restricting
to the conformal mode, but that would involve a trace over all indices, not just those along the brane.
We have not attempted to compute (\ref{dsfin}) in an actual example but can relate this expression to 
CFT computations that have appeared previously in the literature \cite{Perlmutter:2013gua,Hung:2014npa} for the
expansion of the R\'enyi entropy $S_\alpha$ around $\alpha=1$ for a spherical entangling surface. 
We first notice that the first-order variation of entanglement
entropy itself reads
\beq
\delta S_{EE} = \frac{1}{8G_N} \int d^{d-1}\sigma \sqrt{h} h^{ij}\delta h_{ij}\, ,
\eeq
and therefore (\ref{dsfin2}) can be interpreted as a two-point function
\beq \label{aux11a}
C_E = \Delta S^2_{EE} =  \langle \delta S_{EE} \delta S_{EE} \rangle\, ,
\eeq
where we view $\delta h_{ij}$ as a fluctuating bulk field. This clarifies more directly the relation
between the bulk computation and the boundary computation in terms of correlation functions of the modular
Hamiltonian $K$. Indeed, if we interpret the first law of entanglement entropy as an operator statement 
(see e.g. \cite{czech,ownpaper}), so that 
\beq \label{dsfin2b}
\delta S_{EE} \equiv K\, ,
\eeq
then the equivalence between $\langle \delta S_{EE} \delta S_{EE} \rangle$ and $\langle KK\rangle$ becomes manifest.
In this way, we can also much more directly derive bulk expressions for the higher moments of $K$, they
simply become higher order bulk correlators of $\delta S_{EE}$, and in this way perturbatively prove the result of 
\cite{Dong:2016fnf}.

The above computations also nicely illustrate the relation of quantum entanglement
to gravity: we can see that entanglement fluctuations are interpreted as self-gravitation, with the
two-point function (\ref{aux11a}) as the leading order integrated self-gravitation of the bulk surface\footnote{Given the equality (\ref{equalities}) of the
capacity of entanglement, fidelity susceptibility and quantum Fisher information under modular flow, the corresponding bulk interpretations of the
three should also be related. However, on the face of it, proposals for the latter two look quite different \cite{Banerjee:2017qti, Alishahiha:2017cuk, MIyaji:2015mia, Lashkari:2015hha}. It would be interesting (but beyond the scope of this work) to understand in more detail the relation
of the bulk interpretations when (\ref{equalities}) holds.}.

\subsection{Direct CFT computation}\label{directCFT}

As we just described, a more direct CFT computation of the capacity of entanglement and higher moments of the entanglement spectrum
involves the computation of $n$-point functions of the modular Hamiltonian. For spherical regions, the modular
Hamiltonian has a local expression 
\beq
K_{A} = 2\pi \int_{|x|<R}{d}^{d-1}x\frac{R^2-r^2}{2R}T_{tt}(x)
\eeq
and the capacity of entanglement then
involves a double integral of the 2-point function $\bra T_{tt}(x) T_{tt}(x')\ket$.
In the case $d=2$, using a slightly ad hoc way to treat the divergent integral,\footnote{It should be remembered that $T=2\pi T_{zz}$ and ${\bar T}=2\pi T_{{\bar z}{\bar z}}$.}
\bear
C_E = \bra K^2_A\ket_c &=& \int\limits_{-R}^{R} dx\int\limits_{-R}^{R} dx' \frac{(R^2-x^2)(R^2-x'^2)}{4R^2} \frac{c}{(x-x')^4}\\
 &=& \int\limits_{-R(1-\varepsilon)}^{ R(1-\varepsilon)} dx' c\frac{R(R^2-x'^2)}{3(R-x')^2(R+x')^2}\\
 &=& \frac{2c}{3}{\rm ArcTanh}(1-\varepsilon) \\
 &=& \frac{c}{3}\left(\log\frac{2}{\varepsilon}\right) + o(\varepsilon)\\
 &=& \frac{c}{3}\log\frac{L}{a}\, .
\eear
In the above, we used the well-known $ \langle T(z)T(0)\rangle\sim \frac{c}{2z^4}$ formula and also set a UV regulator $\varepsilon= \frac{2a}{L}$ by hand. 
In the first step, we did the $x$ integral assuming the contour had been shifted off the real axis. The second integral is then performed with an explicit cutoff.

One can systematically study higher cumulants and see that they are 
given by the terms that appear in the higher order correlation functions of the energy momentum tensor.
This is explored in quite some detail in \cite{Hung:2014npa}, see also section \ref{Highdalt},
in particular the relation between the
second derivative of the R\'enyi and the coefficient that appear in the three-point function of the
energy momentum tensor is worked out in this paper. This paper also discusses a particular regularization 
of the integrated correlation functions of the type above, but only in a simple example, and proposes to
use dimensional regularization whose connection to standard regulators is obscure.

\section{State dependence of the capacity of entanglement}
\label{Sec4}

The capacity of entanglement depends on the choice of the original state of the system. In this section we study this dependence by considering various
simple examples. 

\subsection{Simple $n$-qubit examples}

\begin{figure}[t] 
\begin{center}
\vspace{-0.cm}
\hspace{-.5cm}
\includegraphics[width=.46\textwidth]{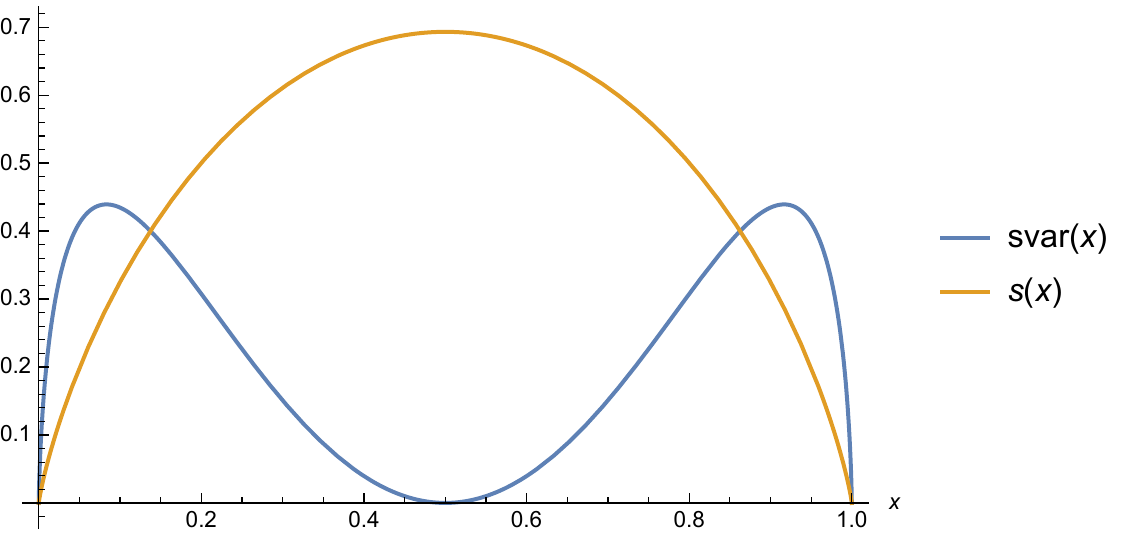}
\end{center}
\vspace{-.3cm}
\caption{\label{fig:ssvar2}
Entanglement entropy and the capacity of entanglement for the state (\ref{staten2}) as a function of $x=\cos^2 (\theta/2)$. 
}
\end{figure}

We begin by studying simple $n$-qubit systems.
Two-qubit systems were discussed in \cite{FeYu}. As a simplest example, consider the states
\beq
 |\theta, \phi \ket = \cos (\theta /2) |10\ket + e^{i\phi} \sin (\theta /2) |01\ket
\eeq
and form a reduced density matrix by tracing over the other spin,
\beq\label{staten2}
\rho_{red} = {\rm diag} (\sin^2 (\theta/2),\cos^2(\theta/2)) \ .
\eeq
The entanglement entropy grows monotonically towards the maximum, but the capacity of entanglement goes to zero  both in the maximally entangled and separable state limits
while reaching a maximum at a partially entangled state with $\cos^2(\theta /2)\approx 0.2885$ (Figure \ref{fig:ssvar2}).\footnote{Ref.  \cite{FeYu} considered a slightly more general example, starting with the states
\beq
 |\psi \ket = a_1|00\ket+a_2|01\ket+a_3|10\ket+a_4|11\ket
\eeq 
with the normalization $\sum^4_{i=1}|a_i|^2=1$. The diagonalized reduced density matrix for one spin takes the same form as above, but \cite{FeYu} wrote it in
terms of the concurrence $c$ of the 2-bit system,
\beq
 \rho_{red} ={\rm diag} (\frac{1+\sqrt{1-c^2}}{2},\frac{1-\sqrt{1-c^2}}{2})
 \ ,
\eeq
where 
\beq
 c= 2|a_1a_4-a_2a_3| \ .
\eeq
They also observed that in general $\Delta S_{EE}^2\neq S_{EE}$, equality holds only at $c=0$ and at special value $c\approx 0.8272$.} 

\begin{figure}[h]
\begin{center}
\includegraphics[width=0.4 \textwidth]{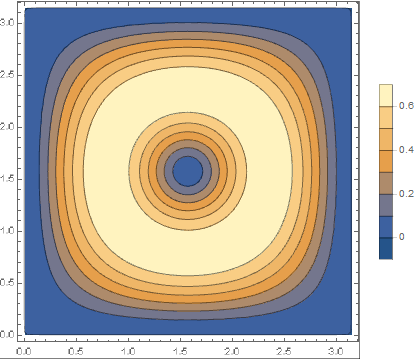}
\hfil
\includegraphics[width=0.4 \textwidth]{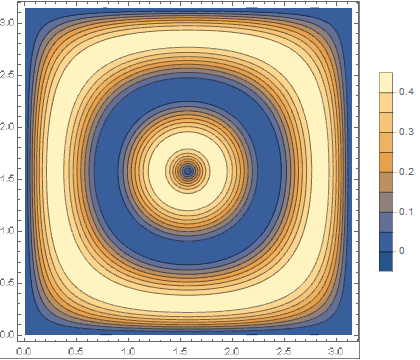}
\\
({\bf A})
\hfil ({\bf B})
 \\
\caption{\label{fig:ssvar3}
Entanglement entropy ({\bf A}) and the capacity of entanglement ({\bf B}) plotted as a function of the angle parameters $\theta,\phi$ of the state (\ref{staten3}).}
\label{QLZ}
\end{center}
\end{figure}

Moving to a three-qubit system, we consider the subspace
\beq\label{staten3}
|\theta,\phi\ket = \cos \theta |001\ket + \sin \theta (\cos \phi |010\ket + \sin \phi |100\ket )
\eeq
and form a reduced density matrix by tracing over one spin. In this case, we again see that the capacity of entanglement goes to zero in the limit of maximal entanglement and separable states, while
reaching a maximum for partially entangled states (Figure \ref{fig:ssvar3}.)

\begin{figure}[t] 
\begin{center}
\vspace{-0.cm}
\hspace{-.5cm}
\includegraphics[width=.46\textwidth]{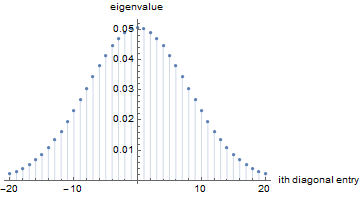}
\end{center}
\vspace{-.3cm}
\caption{\label{fig:entries}
$N=41$ eigenvalues $\lambda_i$ of $\rho$ with a Gaussian profile. The indexing is $i=-20,-19,..,20$. 
}
\end{figure}

As these simple examples illustrate, the entanglement entropy and its fluctuations, characterized by the cumulants have no reason to be equal, except for some special states. 
The capacity of entanglement characterizes the width of the eigenvalue spectrum of the density matrix. We study that by a simple example. Let us assume
that the (reduced) system has $N$ levels, and we have moved to a diagonal basis.  We then assume that the $N$ eigenvalues $\lambda_i$ of the (reduced) density matrix
have a gaussian distribution with variance $\sigma$, {\em i.e.} they form a gaussian vector (an example is shown in Figure \ref{fig:entries}).

\begin{figure}[h]
\begin{center}
\includegraphics[width=0.4 \textwidth]{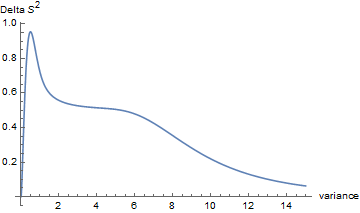}
\hfil
\includegraphics[width=0.4 \textwidth]{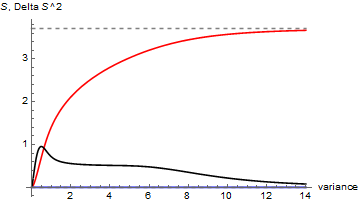}
\\
({\bf A})
\hfil ({\bf B})
 \\
\caption{\label{fig:S-variance}
(Entanglement) entropy and the capacity of entanglement plotted as a function of the variance $\sigma$ of the eigenvalue distribution. The left panel (A) shows the capacity of entanglement
$C_E =\Delta S^2$, the right panel (B) superimposes it (lower curve) with the the entropy (upper curve). The dashed line indicates the maximum entropy.}
\end{center}
\end{figure}

 We can now interpolate from $\sigma=0$, where the density matrix describes a pure state, to $\sigma \rightarrow \infty$ limit giving a maximally mixed state. We then plot the (entanglement) von Neumann entropy and the capacity of entanglement $C_E = \Delta S^2$ as a function
of $\sigma$ in Figure \ref{fig:S-variance}. The left panel (A) shows only the capacity of entanglement to highlight its non-trivial dependence of the spread of eigenvalues; 
the right panel (B) shows it together with the entropy.   

\subsection{Random bipartite entanglement}\label{random}

One could imagine that the origin of entanglement and capacity of entanglement has to do with the fact
that we are projecting a random pure state to a subsystem. It is therefore interesting to compute the
expectation value of the entanglement entropy and capacity of entanglement in such a setting. More precisely, 
we are going to consider random pure states in a Hilbert space of dimension $pq$ of the form ${\cal H}_q 
\otimes {\cal H}_p$, projected to the subsystem with Hilbert space ${\cal H}_p$ of dimension $p$ ($q \geq p$), a review is {\eg} \cite{Majum}.
In this way we will get a probability distribution of reduced density matrices, and we can compute the
expectation value of entanglement entropy and the capacity of entanglement in this ensemble.

The probability distribution for the eigenvalues $\lambda_i$ of the reduced density matrix is \cite{Lloyd:1988cn}
\begin{equation}
P_{p,q}(\lambda_1,\ldots,\lambda_p) = {\cal N}_{p,q} \prod_{1\leq i<j\leq p} (\lambda_i-\lambda_j)^2 \prod_{1\leq k \leq p}
\lambda_k^{q-p} \,\,\delta\left(1-\sum_{i=1}^p \lambda_i\right) .
\end{equation} 
where ${\cal N}_{p,q}$ is a known normalization factor \cite{Zyc}.
To compute the expectation value of the Renyi entropies, we need
to compute the expectation value of $\sum\lambda_i^{\alpha}$ under this probability distribution. 
The relevant integrals can be extracted from the results in \cite{Giraud}. For example, we find that
when $p=q=2$
\be
\langle {\rm tr} \rho^{\alpha} \rangle_{2,2} = {\cal N} \frac{2(\alpha^2+\alpha+2)\Gamma(\alpha+1)}{\Gamma(\alpha+4)}
\ee
from which we obtain
\be
\langle S_{\rm EE}\rangle_{2,2} = \frac{1}{3},\qquad \langle C_{\rm E} \rangle_{2,2} = \frac{13}{36}.
\ee
We therefore see that for $p=q=2$ the entanglement entropy and capacity are very close to each other. 

For arbitrary $p,q$ an exact answer for $\langle S_{\rm EE}\rangle_{p,q}$ was conjectured in \cite{page}
and proven in \cite{proofofpage},
\be
\langle S_{\rm EE}\rangle_{p,q} = \sum_{k=q+1}^{pq} \frac{1}{k} - \frac{p-1}{2 q} \ ,
\ee 
for $p\leq q$. 

Similarly we can compute
\be
\langle {\rm tr} \rho^{\alpha} \rangle_{2,q} = {\cal N} \frac{2 (q-2)!(\alpha^2+\alpha+2q-2)\Gamma(\alpha+q-1)}{\Gamma(\alpha+2q)}
\ee
from which one can obtain for example $\langle S_{\rm EE}\rangle_{2,3} = \frac{9}{20}$
and $\langle C_{\rm E} \rangle_{2,3} = \frac{1169}{3600}$. Numbers quickly become unwieldy, but as we increase
$p$ and $q$ in general the entanglement entropy increases and the capacity divided by entanglement 
decreases. This is consistent with the idea that as $p$ and $q$ increase, the reduced density matrix becomes
more and more maximally mixed. To illustrate, for $p=2$ and $q=100$ we get
$\langle S_{\rm EE} \rangle_{2,100} \simeq 0.686$, close to the maximal value $\log 2 \simeq 0.693$, while
$\langle C_{\rm E} \rangle_{2,100} \simeq 0.015$.

The case $p=q=2$ is the case where entanglement entropy and capacity are nearly equal to each other. 
The ratio $C_{\rm E}/S_{\rm EE}$ decreases even when we keep $p=q$ and increase $p$ and $q$. 
Already for $p=q=3$ where $S_{\rm EE}=1669/2520$ and $C_{\rm E}=2898541/6350400$ the ratio has decreased to
$C_{E}/S_{EE}\simeq 0.689$.  As we briefly review below, this can be proved rigorously be studying a the
limit of large $p,q$ using random matrix theory \cite{pagelargen,Chen:2017yzn}.

The main observation of these studies is that a possible explanation for the approximate equality of $S_{EE}$ and $C_{E}$ in a system is that most
of the entanglement is being carried by randomly entangled pairs of qubits (the case $p=q=2$). 

\subsubsection{Random pure states and Wishart-Laguerre random matrices}

We may also study the reduced density matrices of random pure states by using their connection
to the Wishart-Laguerre ensemble of random matrices. We are not going to perform a mathematically rigorous analysis here (along the lines of \cite{pagelargen}), but just quickly study whether
the ratio $C_{\rm E}/S_{\rm EE}$ decreases to zero in the limits $p=q\rightarrow \infty$, and $p,q\rightarrow \infty$ with $p/q\rightarrow 0$,
 following the approach in \cite{Chen:2017yzn}.  
In \cite{Chen:2017yzn}, the reduced density matrix
was written in the form
\beq
\rho = \frac{YY^\dagger}{\Tr(YY^\dagger)} \ ,
\eeq
where $W\equiv YY^\dagger$ is a $p\times p$ random matrix belonging to the  $\beta =2$
Wishart-Laguerre ensemble. In the limit of large $p,q$ with $\alpha = p/q$ fixed,
one may replace $\Tr(YY^\dagger )\rightarrow p$, so that the eigenvalues $\lambda_i$ of $\rho$
are related to the eigenvalues $\mu_i$ of the Wishart matrix $W$ by a simple rescaling
\beq
  \lambda_i = \frac{\mu_i}{p} \ .
\eeq
This gives a quick way of computing the ensemble average of the R\'enyi entropy, by using
the average spectral density of the Wishart ensemble.
We start with
\beq
  k(s) = \Tr (\rho^s) = \sum_i \lambda^s_i \approx \frac{1}{p^s} \int d\mu \sum_i \delta (\mu-\mu_i) \mu^s
\eeq
where 
\beq
 \sum_i \delta (\mu - \mu_i) \equiv \hat{\nu}(\mu)
\eeq
is the spectral density of the Wishart matrix. Its ensemble average $\bra \hat{\nu}(\mu)\ket$
is given by the Marchenko-Pastur distribution \cite{MP} $\bar{n}(\mu)$,
\beq
   \bra \hat{\nu}(\mu)\ket = p \bar{n}(\mu) = \frac{p}{2\pi \alpha \mu}
\sqrt{(\mu - \alpha_-)(\alpha_+ - \mu) } \ ,
\eeq
where $\alpha_- \leq \mu \leq \alpha_+$, with $\alpha_\pm = (1\pm \sqrt{\alpha})^2$, $\alpha = p/q$. The ensemble average  $\bra k(s)\ket$ thus becomes
\begin{eqnarray}\label{ksrandom}
&& \bra k(s) \ket = \bra \Tr \rho^s \ket = \frac{1}{p^{s-1}} \int d\mu~\bar{n}(\mu) \mu^s \\ \nonumber
&& = p^{1-s} \frac{1}{2\pi\alpha} \int^{\alpha_+-\alpha_-}_0 dy \sqrt{y(\alpha_+-\alpha_--y)}(y+\alpha_-)^{s-1} \\ \nonumber
&& =\left(\frac{ p}{ \alpha_-}\right)^{1-s}~\mbox{}_2F_1 \left(1-s,\frac{3}{2},3,\frac{-4\sqrt{\alpha}}{\alpha_-}\right) \ .
\end{eqnarray}
The general result from (\ref{ksrandom}) for the entanglement entropy and the capacity of entanglement is  then
\begin{eqnarray}
 && \bra S_{EE} \ket = \log \frac{p}{\alpha_-} + G^{(1)}_a \left(0,\frac{3}{2},3,\frac{-4\sqrt{\alpha}}{\alpha_-}\right) \\ \nonumber
&& \bra C_E \ket = G^{(2)}_a\left( 0,\frac{3}{2},3,\frac{-4\sqrt{\alpha}}{\alpha_-}\right) -  \left(G^{(1)}_a \left(0,\frac{3}{2},3,\frac{-4\sqrt{\alpha}}{\alpha_-}\right)\right)^2
 \end{eqnarray}
where\footnote{Properties of these derivatives of the Hypergeometric function can be found in \cite{Ancarani}. They are also built into the Mathematica program. }
\beq
  G^{(n)}_a (a,b,c,z)= \frac{\partial^n}{\partial a^n} \mbox{}_2F_1(a,b,c,z) \ .
\eeq
In the limit $\alpha = p/q \rightarrow 0$, one obtains
\beq
   \frac{\bra C_E \ket}{\bra S_{EE}\ket} \approx \frac{\frac{p}{q}}{\log p -\frac{p}{2q}} \rightarrow 0 \ ,
\eeq
confirming the numerical guess.
In the limit $p=q$ (\ref{ksrandom}) gives
\begin{eqnarray}
&&\bra k(s) \ket = p^{1-s} \frac{1}{2\pi} \int^4_0 dy\sqrt{y(4-y)}y^{s-1} \\ \nonumber
&& = \frac{4}{\sqrt{\pi}}\left(\frac{p}{4}\right)^{1-s}  \frac{\Gamma(s+\frac{1}{2})}{\Gamma (s+2)} \ ,
\end{eqnarray}
in agreement with Eqn. (12) in \cite{pagelargen}, 
from which we obtain
\beq
  \bra S_{EE} \ket \approx \log p -\frac{1}{2}\ ; \ \bra C_E \ket \approx \frac{\pi^2}{3}-\frac{11}{4} \approx 0.5399
\eeq
also confirming the numerical guess $\bra C_E\ket/\bra S_{EE}\ket \rightarrow 0$ in the limit $p=q\rightarrow \infty$.

We therefore see that a possible explanation for approximate equality of $S_{EE}$ and $C_{E}$ in a system is that most
of the entanglement is being carried by randomly entangled pairs of qubits.

\section{The capacity of entanglement in quantum field theories and universality}
\label{Sec5}

\subsection{1+1 dimensional CFTs}

We now move to consider conformal field theories in various dimensions. In particular, in 1+1 dimensions we can make use of very general analytical results for the R\'enyi entropies.
R\'enyi entropies for generic CFTs  have been computed in a variety of cases: for an initial vacuum state for infinite and finite systems, for thermal 
backgrounds for finite and infinite systems, and for non-equilibrium dynamics involving different quench protocols, notably global and local quenches\footnote{In Section \ref{Sec2} we discussed the relation of the R\'enyi entropies and the entanglement spectrum. A recent
paper \cite{Wen:2018svb} studies the time evolution of the entanglement spectrum and the entanglement Hamiltonian after a quench.}. A recent review is {\em e.g.}
\cite{Calabrese:2016xau}.
The computations lead to
\beq\label{CFTRenyi}
\Tr(\rho_A^\alpha) = c_\alpha  e^{-\frac{c}{12}\left(\alpha-\frac{1}{\alpha}\right)W_A},
\eeq
where $W_A$ is a function of the size of the subsystem (and time, for quenched systems) but independent of $\alpha$, and $c_\alpha$ with $c_1=1$ are model specific constants that do not depend on the size $l$. The R\'enyi entropies $S_\alpha (A)$ are then 
\begin{eqnarray}
 &&  S_\alpha (A) = \frac{c}{12} \left(1 +\frac{1}{\alpha}\right) W_A + (1-\alpha)^{-1} \log c_\alpha %
\end{eqnarray}
From the cumulant generating function $\tilde{k}(\alpha)=(1-\alpha )S_\alpha$ , we immediately obtain the general result for the capacity of entanglement,
\beq\label{CFT2variance}
C_E =  S_{EE}+c'_1-(c'_1)^2+c''_1 = \frac{c}{6} W_A + c'_1 + (c'_1)^2 + c''_1\ ,
\eeq
and in particular, for large $W_A$, the last three terms on the right hand side are negligible. 
Explicit results for $W_A$ for various different situations are listed in \cite{Cardy:2016fqc}. For time independent cases with $A$ a finite interval of length $l$,
\beq\label{cases1}
W_ A = \begin{cases}	2\log \left(\frac{l}{\epsilon}\right) & {\rm infinite\ system},\,\,\, T= 0\\
 			2\log \left(\frac{L}{\pi \epsilon}\sin\frac{l \pi}{L}\right)  & {\rm finite\ system,\ size\ } L,\,\,\, T= 0\\
			2\log \left(\frac{\beta}{\pi \epsilon}\sinh\frac{l \pi}{\beta} \right) & {\rm infinite\ system} ,\,\,\, T> 0 \ .\end{cases}
\eeq 
For time-dependent cases with two semi-infite intervals, some known results are
\beq
W_ A = \begin{cases}	\log \left(\frac{\beta}{\pi \epsilon}\cosh (2\pi t/\beta )\right) & {\rm global\ quench\ at}\,\, t= 0\\
 			\log \left(\frac{t^2 + \lambda^2 }{\epsilon \lambda /2}\right)  & {\rm local\ quench\ at} \,\, t= 0 \ , \end{cases}
\eeq
with quench parameters $\beta$ (the resulting inverse temperature) and $\lambda$. 

It may be somewhat surprising that the leading order result $C_E = S_{EE}$ holds in all of the above cases, even for the quenched nonequilibrium systems throughout
their time evolution. The entangled states are rather special, prepared by using conformal mappings.
Also, the result tells us that after the quench, the growth rate of the capacity of entanglement is the same as that of the 
entanglement entropy. The growth of the entanglement entropy is consistent with the interpretation where the spreading of entanglement is carried by quasiparticle pairs created at the quench and propagating ballistically through the system at the speed of light. It is not obvious at all that
this picture should imply the same growth rate for the capacity of entanglement. Consider for example the following toy model.
Suppose we quench a system and afterwards entanglement builds up because pairs of perfectly entangled quasiparticles
are created. The reduced density matrix of a subsystem will then roughly consist of
the original reduced density matrix combined with say $n(t)$ qubits which are pairwise perfectly
entangled with $n(t)$ qubits outside the subsystem. The number of such qubits will be
time dependent. We will model the reduced density matrix as
\beq
\rho (t) = \rho_{UV} \otimes \left(\frac{1}{2^{n(t)}}\mathbb{I}_{2^{n(t)}\times2^{n(t)}}\right)\otimes |\psi (t)\rangle\langle \psi(t)|
\eeq
where $\rho_{UV}$ represents the very UV degrees of freedom which were entangled, and remain entangled across the boundary of the region. The pure state $\psi (t)$ represents
the reservoir of states from which qubits are extracted as more and more entangled
quasiparticles become relevant. This pure state lives in some Hilbert space of dimension $D-2^{n(t)}$ with a very large $D$. Details of the reservoir pure state are irrelevant. It is
straightforward to compute the R\'enyi entropies for this density matrix,
\beq
\frac{1}{1-\alpha}\log\Tr (\rho(t)^\alpha) = S^\alpha_{UV} + n(t) \log 2 
\eeq
We then deduce that
EE grows as a function of $t$,
\beq
  S_{EE} = S^{UV}_{EE} + n(t) \ln 2 \ ,
\eeq
where the function $n(t)$ can be adjusted to the quasiballistic interpretation (with the characteristic long linear growth regime), but the capacity of entanglement
stays constant 
\beq
C_{E} = \Delta S^2_{UV}
\eeq
receiving contributions only from the UV, not from the created quasiparticles.
Therefore if the capacity of entanglement and
EE grow identically after the quench, it is in tension with the simple ballistic  propagation of created pairwise perfectly entangled quasiparticles which
leads to the above model. 

 Rather than expecting the quasiparticles to be pairwise perfectly entangled, we could expect them to be created
in random pure states with some statistical distribution. In Section \ref{random} we studied random pure states with bipartite entanglement between degrees of freedom inside and outside the subsystem. There we saw that it was indeed possible for the capacity of entanglement to be approximately equal to the entanglement entropy, for ensemble averages of randomly entangled  pairs of qubits.
On the other hand, in Section \ref{Nfluct} we saw that such an equality also arises  (in the large $N$ limit)  from a more detailed connection between particle fluctuations into a subregion and its entropy. We conclude that the intuitive interpretation of entanglement being carried by ballistic propagation quasiparticles needs to be refined with correct statistics for the
distribution of entanglement.

Finally, we make some brief comments on the two interval case. Let the two intervals be $A=[u_1,v_1]$ and $B=[u_2,v_2]$. The R\'enyi entropy for the two interval case has the general form \cite{mutuinfo}
\beq
  \Tr \rho^\alpha_{A\cup B} = c_\alpha \left(\frac{(u_2-u_1)(v_2-v_1)}{|(v_2-u_1)(v_2-u_2)(v_1-u_1)(v_1-u_2)|} \right)^{(c/6)(\alpha - \frac{1}{\alpha})} {\cal F}_\alpha (\eta)
\eeq
where $c_{\alpha}$ is a non-universal constant and ${\cal F}_{\alpha}$ is a non-universal function of the cross ratio  $\eta$. This non-universal function makes the analysis more complicated,
so here we consider only the case ${\cal F}_{\alpha}=1$, such as in the massless fermion theory \cite{Casini:2005rm,Casini:2008wt,Casini:2009vk}. Then the
R\'enyi entropy takes the same form as (\ref{CFTRenyi}), and we readily obtain a similar relationship for the leading terms as in (\ref{CFT2variance}),
\beq
C_E (A\cup B) = S _{EE}(A\cup B) \ .
\eeq

\subsection{An alternative calculation in higher dimensions}
\label{Highdalt}

In subsection \ref{directCFT} we discussed an alternative calculation to compute the capacity of entanglement for a 1+1 dimensional
conformal field theory. In this section we discuss its generalization to higher dimensional CFTs.

Suppose we consider the entanglement of a ball. Inserting the explicit form  of the modular Hamiltonian, we obtain a suitably integrated two-point 
function of the energy-momentum tensor, in agreement with (\ref{aux11a}) and (\ref{dsfin2b}). 
Such types of integrated correlation functions were studied in \cite{Perlmutter:2013gua,Hung:2014npa}. The idea
of the computation is as follows. A ball in a constant time slice is mapped to itself under a subgroup
$SO(1,d-1)$ of the conformal group, and the double integral is invariant under the action of this
conformal group. We can use this symmetry to fix one of the points at the origin, leaving us with a 
single integral over the ball. This integral is still divergent, but because the modular Hamiltonian
involves the integral of a conserved current over a fixed timeslice, we can evaluate it on any suitable
time slice, and in particular we can move the slice a little bit so as to avoid the origin. What is then
left is a finite integral, which we still need to multiply by the volume of the gauge group that we
fixed. This volume is the volume of $SO(1,d-1)/SO(d-1)$, which is the Euclidean hyperbolic space, because 
the origin is left fixed by a $SO(d-1)$ subgroup of $SO(1,d-1)$. This hyperbolic space is the same
space that appears at the boundary of AdS if we map the Rindler wedge associated to the ball to a 
hyperbolic black hole as in \cite{Casini:2011kv}. 

It is clear from the above that the final result will only depend on the coefficient $C_T$ which
appears in the two-point function of two energy-momentum tensors
\beq
\langle T(x)T(0)\rangle \sim \frac{C_T}{x^{2d}}.
\eeq
The relevant computation of the integrated two-point function was studied in detail in 
\cite{Perlmutter:2013gua}
for arbitrary CFT's, and equation (1.3) therein reads
\beq \label{perl}
S'_{q=1} = -{\rm Vol}\,{\mathbb H}^{d-1} \cdot \frac{\pi^{d/2+1} \Gamma(d/2) (d-1)}{(d+1)!} C_T
\eeq
where the volume of hyperbolic space appears as we just explained, and we can interpret the result as
the capacity of entanglement by $C_E =\Delta S^2_{EE} = -2 S'_{q=1}.$
The result is indeed proportional to the coefficient $C_T$ that appears in the two-point function
of two energy-momentum tensors.
For later use, we write the result in a different form.
As also explained in \cite{Perlmutter:2013gua}, the $C_T$ above should not be confused with the coefficient $c$ 
appearing in the $d=4$ trace anomaly of a CFT or the ${\tilde{C}}_T$ used in \cite{Hung:2011nu} in their expression for the R\'enyi entropies. 
They are proportional to each other, 
\beq
{\tilde{C}}_T=\frac{\pi^d(d-1)}{(d+1)!}C_T
\eeq
such that ${\tilde{C}}_T|_{d=4}=c$. Thus, from (\ref{perl}), we arrive the expression
\beq
 C_{E} =  {\tilde{C}}_T \frac{2 \Gamma (d/2)}{\pi^{d/2-1}}{\rm Vol}\,{\mathbb H}^{d-1} \ .
\eeq

This expression is similar to that of the entanglement entropy. When computing the entanglement
entropy via the conformal mapping to hyperbolic space, the original computation in \cite{Casini:2011kv} obtained the expression
\beq
S_{EE}=a_d^*\frac{2\Gamma\left(d/2\right)}{\pi^{d/2-1}}{\rm Vol}\,{\mathbb H}^{d-1}\,,
\eeq
where $a_d^*$ is a dimensionless constant, a central charge. In even dimensional spacetimes, this equals the coefficient of the A-type trace anomaly of the CFT. In odd dimensional spacetimes, $a_d^*\propto \log Z_{S^d}$, i.e. the logarithm of the partition function of the CFT on a unit sphere.

The regulated hyperbolic volume is given explicitly in (3.1) in \cite{Perlmutter:2013gua} and is also easy to compute 
independently. The expression is sensitive to the choice of the regularization scheme, but even and odd $d$ have a 
universal coefficient for the logarithmic and constant term, respectively. They are
\beq
{\rm Vol}\,{\mathbb H}^{d-1}=\frac{\pi^{d/2}}{\Gamma(d/2)}\begin{cases}(-)^{d/2-1}2\pi^{-1}\log(R/\varepsilon)& $d$\,\mathrm{even}\\
(-)^{(d-1)/2}& $d$\,\mathrm{odd}\end{cases}.
\eeq
To conclude, the above computations suggest a simple universal ratio for the entanglement entropy and the capacity of entanglement,
\beq\label{endratio}
\frac{C_{E}}{S_{EE}} = \frac{{\tilde C}_T}{a^*_d} \ .
\eeq
We discuss universality in more detail in Section \ref{unisec}.

\subsection{The entanglement spectrum}
\label{enspec}

The capacity of entanglement captures a particular feature of the full entanglement spectrum. Here we briefly
review some aspects of the full entanglement spectrum and comment on the implications for the capacity of
entanglement. 
 
We start with the two-dimensional case where the entanglement spectrum can be explicitly obtained in several cases.
In particular, from (\ref{CFTRenyi}) one can in principle extract the entanglement spectrum for 1+1 dimensional
CFT's using an inverse Laplace transform. Of course, the precise answer will depend on the detailed structure 
of $c_{\alpha}$, but it is instructive to show the spectrum in 
an explicit example. For $c_{\alpha}=1/\alpha$, the inverse Laplace transform leads to the following result
for the density of eigenvalues of the reduced density matrix \cite{CalLef,Alba:2017bgn}
\beq
\rho(\lambda) = \theta(-\log\lambda - c W_A/12) \lambda^{-1} I_0(2\sqrt{c W_A(-\log\lambda-W_A)/12} )
\eeq
and one can verify explicitly that indeed
\beq
\int_0^1 \rho(\lambda) \lambda^{\alpha} = \alpha^{-1}  e^{-\frac{c}{12}\left(\alpha-\frac{1}{\alpha}\right)W_A} .
\eeq
It is perhaps more instructive to rewrite this in terms of an energy spectrum with $\lambda=e^{-E}$. If we denote
\beq 
E_c=\frac{c}{12} W_A
\eeq
then
\beq \label{yy0}
\rho(E) = \theta(E-E_c) I_0(2\sqrt{E_c(E-E_c)} )
\eeq
and indeed
\beq 
\int_0^{\infty} dE \rho(E) e^{-\alpha E} = \alpha^{-1}  e^{-\frac{c}{12}\left(\alpha-\frac{1}{\alpha}\right)W_A}.
\eeq
It is interesting to see that the spectrum cuts off at a value $E_c$ which is determined by the UV cutoff that one
needs to introduce in order to regulate the divergences in the computation of the R\'enyi entropies. One is 
almost tempted to view $E_c$ as some sort of analog of the Casimir energy. As we remove
the cutoff, all eigenvalues of the density matrix are located in an ever smaller interval starting at $\lambda=0$. 
In the above, we only considered the leading term of the R\'enyi entropy in terms of a cutoff, one can in principle
do more precise computations taking the exact cutoff dependence into account (which involves a choice of boundary
state) \cite{CLR,Alba:2017bgn}, but that is beyond the scope of this paper.

Ultimately, in 1+1 dimensions, (\ref{yy0}) follows from conformal invariance, but in a rather indirect way. It would
be desirable to have a more direct understanding of the connection between conformal invariance and
the various features of (\ref{yy0}), as that might shed further light on the relation between 
$C_{E}$ and $S_{EE}$ in higher dimensions. 

Expressions where $c_{\alpha}$ is some other integer power of $\alpha$ can in principle be obtained
from (\ref{yy0}) by differentiating or integrating with respect to $E$. The feature of (\ref{yy0}) which
does not depend much on these details is its large $E$ behavior. For large $z$, the asymptotic behavior of
$I_{\nu}(z)$ is $I_{\nu}(z)\sim e^z/\sqrt{2\pi z}$. Keeping only the exponential we therefore see that
$\rho(E)$ behaves as
\beq \label{densstates}
\rho(E) \sim \exp\left( 2\sqrt{E_c(E-E_c)} \right), \qquad E\rightarrow \infty
\eeq
Another interpretation of this expression is that it is the density of states for a 1+1 dimensional CFT on
the plane. In higher dimensions, it maps to the density of states on the hyperbolic plane, but the 
one-dimensional hyperbolic plane is just the ordinary line. 
Actually, according to \cite{Casini:2011kv} after a change of coordinates the causal diamond
attached to the spatial interval $r\in[-l/2,l/2]$ maps under the change of coordinates 
\beq
\label{delcoord}
t\pm r = \tanh \left(\frac{1}{2} \left(\frac{\tau}{R}\pm u\right)\right)
\eeq
to
\beq
ds^2 = \frac{-d\tau^2 + R^2 du^2}{(\cosh u + \cosh(\tau/R) )^2}
\eeq
where, for an interval of length $l$, $R=l/2$. Putting a UV cutoff at $r=\pm(R-\epsilon)$ corresponds under
this change of coordinates to $u\simeq \pm \log(l/\epsilon)$ for small $\epsilon$. The length of the $u$-interval
is therefore precisely what we called $W_A$.

We therefore need to compute the density of states on the plane for a box of size $W_A$ and relate the energy
on the plane to the energy on the causal diamond. The change of coordinates (\ref{delcoord}) has a 
Schwarzian derivative of $-1/2$. To get the full shift in the energy we need to integrate this over the spatial
box. Therefore the full shift in the left- or right-moving energy is $c W_A/24$. This leads to a shift in
the total energy of the form
\beq
E_{\rm plane} \sim E_{\rm diamond} -\frac{c}{12}W_A
\eeq
Moreover, the energy spectrum for a box of size $W_A$ is quantized in units of $\sim 2\pi/W_A$, and there is an
extra factor of $2\pi$ relating the energy to $L_0$, so altogether what one should use in the Cardy formula is 
\beq
L_0^{\rm plane}\sim \frac{W_A}{(2\pi)^2} (E_{\rm diamond} -\frac{c}{12}W_A)
\eeq
which explains the form of (\ref{densstates}) from the usual Cardy formula.

In higher dimensions, we can study features of the entanglement spectrum whenever the CFT has a holographic dual,
since the R\'enyi entropies can then be obtained from the thermodynamics of hyperbolic black holes \cite{Hung:2011nu}.
The density of states will now obey
\beq \label{denss}
\int_0^{\infty} dE \rho(E) e^{-\alpha E} =   e^{-f(\alpha)}
\eeq
where 
\beq
f(\alpha) = \alpha S_{EE} \left( 1 - \frac{x_{\alpha}^d}{2} - \frac{x_{\alpha}^{d-2}}{2} \right)
\eeq
with 
\beq
x_{\alpha} = \frac{1+\sqrt{1+ \alpha^2 d (d-2)} }{\alpha d} .
\eeq
We notice that as $\alpha\rightarrow\infty$, $f(\alpha)\sim E_c \alpha$, where 
\beq
E_c = S_{EE} \left(1 - \left(\frac{d-2}{d}\right)^{\frac{d}{2}} - \left(\frac{d-2}{d}\right)^{\frac{d}{2}-1} 
\right)
\eeq
which implies in particular that $\rho(E)=0$ for $E<E_c$. It is tempting to interpret $E_c$ once more as some
sort of Casimir energy. The high-energy behavior of $\rho$ is determined by the behavior of $f(\alpha)$ as
$\alpha\rightarrow 0$ which is $f(\alpha)\sim c_d S_{EE} \alpha^{1-d}$ from which it follows that for large
$E$
\beq \label{car2}
\log \rho(E) \sim c_d' E_c^{1/d} (E-E_c)^{(d-1)/d}
\eeq
where $c_d'$ is some constant which only depends on the dimension $d$. The scaling with $E$ is exactly what
one expects generically for a CFT at high energy/temperature, this simply follows from extensivity of the 
free energy. Corrections to (\ref{car2}) can be obtained from subleading terms in the expansion near $\alpha=0$. The
first subleading term scales as $\alpha^{3-d}$ which gives rise to corrections to (\ref{car2}) of the form
\beq \label{car3}
\log \rho(E)_{\rm subleading} \sim c_d'' E_c^{3/d} (E-E_c)^{(d-3)/d}
\eeq

It is not clear whether these results also hold for more general CFT's without a holographic dual. We also
observe that the capacity of entanglement is related to the behavior of $f(\alpha)$ near $\alpha=1$, whereas
the cutoff energy $E_c$ and the high-energy behavior are related to the large and small $\alpha$ behavior of
$f(\alpha)$. A priori, these are unrelated to each other, and therefore the capacity of entanglement does not
in general make any prediction for the behavior of the density of states at either high or low energy. If the
theory has a holographic dual we know the explicit form of $f(\alpha)$, which only depends on the dimension
$d$, and there are some simple factors of order unity which relate the behavior for small $\alpha$, $\alpha$ of
order $1$, and large $\alpha$. We also confirm once more that $C_E=S_{EE}$ in because $C_E=-f''(1)=S_{EE}=f'(1)$ by
explicit computation. 

We notice that in the above cases not just the capacity of entanglement, but all R\'enyi entropies obey an area law.
This implies that the function $f(\alpha)$ in (\ref{denss}) has an area law and scales as $a^{2-d}$ where $a$ is
a short-distance UV cutoff. This translates into the following leading cutoff dependence of the density of states
\beq
\log\rho(E) \sim \frac{1}{a^{d-2}} \log\hat{\rho} (E a^{d-2}) + \ldots
\eeq
where $\hat{\rho}$ does not depend on the cutoff. This is precisely the scaling one would get from a local Rindler
point of view. Therefore, the area law for capacity of entanglement (and more generally for the R\'enyi entropies)
seems to arise from the fact that these quantities are dominated by UV degrees of freedom localized near the entangling
surface, which in turn can be well approximated by local Rindler modes.

\subsection{The capacity of entanglement in holographic systems}
\label{Holosec}

Entanglement R\'enyi entropy was considered for systems with a holographic dual with a black hole in \cite{Hung:2011nu}. Their method was to relate R\'enyi entropies of spheres to thermal entropies in hyperbolic spacetime at a range of temperatures. This method is applicable only for systems with bulk duals that admit black hole solutions at various temperatures. After obtaining the R\'enyi entropies, the computation of entanglement entropy and the capacity of entanglement are straightforward. All the holographic R\'enyi entropies are proportional to the volume of hyperbolic space, which contains all the divergent terms. As discussed in \cite{Hung:2011nu}, the ratio $S_\alpha /S_1$ of the R\'enyi entropies then yields universal information characterizing the dual CFTs.
In similar spirit, we choose to study the ratio of the entanglement entropy and the capacity of entanglement, another universal constant (at least for spherical entangling surfaces).

In holographic duals of Einstein gravity, the entanglement entropy and the capacity of entanglement are equal,
\beq
S_{EE} = 2\pi\left(\frac{\tilde{L}}{l_p}\right)^{d-1} {\rm Vol}\,{\mathbb H}^{d-1} = C_E\ ,
\eeq
where ${\rm Vol}\,{\mathbb H}^{d-1}$ denotes the volume of the hyperbolic space. Here $\tilde{L}$ is the curvature scale of the dual AdS spacetime. In Einstein gravity, it equals the curvature scale in the cosmological constant term but it's not the case for higher curvature theories.

In Gauss-Bonnet gravity, the ratio becomes more interesting. First, for $d=4$, the bulk and boundary theories have two central charges, $a$ and $c$, that appear in the boundary Weyl anomaly. The entanglement entropy and the capacity of entanglement are
\beq
S_{{EE}} = \frac{2a}{\pi}{\rm Vol}\,{\mathbb H}^{d-1}, \quad C_E = \frac{2c}{\pi}{\rm Vol}\,{\mathbb H}^{d-1} \  ,
\eeq
so that in $d=4$ the gravity calculation indeed produces the same ratio (\ref{endratio}) as the field theory calculation.
A priori, one might not have expected any constraints for the ratio of the two. However, the ratio of the CFT central charges is restricted by the ``conformal
collider bounds'' proposed in \cite{Hofman:2008ar} and proven in \cite{Hofman:2016awc}, implying a bound 
\beq\label{SSratio}
\frac{18}{31} \leq \frac{C_E}{S_{EE}} = \frac{c}{a} \leq 3 \ ,
\eeq
for unitary 4d CFTs with gravity duals. The central charges $a$ and $c$ are known explicitly for some field theories.
It is interesting to compute the ratio even when the theories are not holographic -- one may ask if the holographic "prediction" (\ref{SSratio})
still holds.
 For example, for conformal free scalar field theory and massless free Dirac fermion theory the ratios of $c$ and $a$ are \cite{Birrell:1982ix}
\beq
\frac{c_{{\rm scalar}}}{a_{{\rm scalar}}} = 3, \quad \frac{c_{{\rm fermi}}}{a_{{\rm fermi}}} = \frac{18}{11} \ .
\eeq

For general $d\geq4$ the theory is parametrized with $\lambda$ that is the coefficient for the 4-dimensional Euler density term \cite{Hung:2011nu},
\beq
I = \frac{1}{2l_p^{d-1}}\int
d^{d+1}x\sqrt{-g}\left[\frac{d(d-1)}{L^2}+R+\frac{\lambda
L^2}{(d-2)(d-3)}\chi_4\right] \ .
\eeq
In this case one finds 
\bear
S_{{EE}} &=& 2\pi \left(1+\frac{(d-1)(-1+\sqrt{1-4\lambda})}{d-3}\right)\left(\frac{\tilde{L}}{l_p}\right)^{d-1}{\rm Vol}\,{\mathbb H}^{d-1}, \\
 C_E &=&  2\pi\sqrt{1-4\lambda} \left(\frac{\tilde{L}}{l_p}\right)^{d-1}{\rm Vol}\,{\mathbb H}^{d-1} 
\eear
so that the ratio $C_E/S_{EE}$
is again a universal $\lambda$- and $d$-dependent constant. Here, $\tilde{L}$ is proportional to the $L$ in the action, $\tilde{L}=L\sqrt{2\lambda}/(\sqrt{1-\sqrt{1-4\lambda}})$.

\subsection{Violation of the area law}
It has been known for some time that the area law of entanglement entropy is violated in some systems, we point out that the same is true for the capacity of entanglement. 
Apart from the logarithmic scaling in $1+1$ dimensional conformal field theories, the most common example is provided by critical systems with a finite Fermi surface such as free fermions or Fermi liquids \cite{Wolf:2006zzb,Gioev:2006zz,Swingle:2009bf}. The R\'enyi entanglement entropy of these systems has been computed to the leading order in \cite{Swingle:2010jz}
\beq
S_{\alpha} = \left(1+\frac{1}{\alpha}\right)\frac{1}{(2\pi)}\frac{1}{24}\int_k\int_x {\rm d}A_k{\rm d}A_x|n_x\cdot n_k|\log\left(\frac{L}{\varepsilon}\right)
\eeq
where the integrals are taken over the entangling surface and Fermi surface, the $n$'s are their normal unit vectors and $L$ corresponds to the effective system length.
The capacity of entanglement again tracks the entanglement entropy, with $C_E = S_{EE}$,
 both scaling as $L^{d-1}\log(L)$.

\subsection{On the universality of the ratio of entanglement entropy and the capacity of entanglement}
\label{unisec}

We have already seen in many field theories that the leading order of divergence of both the entanglement entropy and the capacity of entanglement are the same. It's interesting to ask if the ratio of the leading terms, computed using the same regularization scheme, is universal or if it is scheme dependent. At first thought, there is no immediate reason why there would not be some universality. After all, they have the same power dependence on the regularization parameter. However, the choice of the regularization scheme could contain some hidden dependence on the quantity we are computing, spoiling the universality. We demonstrate this below by two cases, free massless scalar fields and fermions.

First, we consider the free massless scalar field theory with both $d=3$ and $d=4$. The Hamiltonian of the theory is
\beq
H = \frac{1}{2}\int_{\mathbb{R}^{d}}~d^dx \left[({ \pi})^2(x)+(\nabla{ \phi})^2(x)
\right] \ ,
\eeq
with the standard commutation relations. The entangling surface is a sphere. As the first way of regularizing the theory, we expand the scalar field $\phi$ and its conjugate momentum field $\pi$  in Fourier modes for $d=3$ (and in spherical harmonics for $d=4$), and discretize the remaining radial integral to a sum. The entanglement entropy and variance can be expressed in terms of correlation matrices inside the sphere. This method of computation was originally used in \cite{Srednicki:1993im} and a more detailed explanation can be found in \cite{Casini:2009sr}.

Using these methods, we performed a quick numerical computation of the entanglement entropy and the capacity of entanglement at various radii and found that
\beq\label{scalarresults}
\frac{C_E}{S_{EE}} \approx \begin{cases} 3 & d=3 \\ 5& d=4 \end{cases}\,.
\eeq
These ratios are only approximate. The R\'enyi entropy of this model in the four dimensional case was considered to greater accuracy in \cite{Kim:2014nza} using the same regularization scheme and can be used to confirm our approximate ratio.

As the second alternative regularization scheme we consider the heat kernel method. In this way,
the R\' enyi entropy of the free massless scalar field theory for a spherical region was also computed analytically in \cite{Solodukhin:2011gn}.  
The effective action, $W_n=-\log Z_n$, of the $n$-sheeted spacetime is
\beq
W_{n}=-\frac{1}{2}\int_{\varepsilon^2}^{\infty}\frac{{\rm d}s}{s}\Tr K(s,n),
\eeq
where the regularization is done by limiting the lower limit of the integral. The heat kernel is defined as
\beq
K(s,n,X,X') = \,_n\langle X|\rm{e}^{-s {\mathcal D}} |X'\rangle_n,
\eeq
where $\mathcal{D}$ is the field operator of the theory. The trace itself is
\beq
\Tr K(s,n) = \frac{1}{(4 \pi s)^{d/2}}\left(nV+\pi\frac{(1-n^2)}{3n}s A(\Sigma)\right).
\eeq
With these, it can be seen that the ratio 
\beq
\frac{C_E}{S_{EE}} = 1
\eeq
for the leading terms, exactly, for all $d$. This is an obvious disagreement with the previous result (\ref{scalarresults}). We conclude that at least for the free massless scalar field theory, the ratio of the leading terms is scheme dependent. %

While the universality fails for generic field theories, we may restrict our consideration to conformal field theories. An alternative regularization scheme valid for all R\'enyi entropies of conformal field theories is provided by the holographic computation technique used in \cite{Casini:2011kv,Hung:2011nu} and discussed above.  We can test universality by the 
following criterion.
We separately expand
the entanglement entropy $S_{EE}$ and the heat capacity $C_E$, and then compare one by one the ratios of the 
leading terms and the ratios of subleading terms at the same order in the expansion.   
If universality holds, the ratios of respective terms in the expansions should all be the same in all regularization schemes.
Conversely, if we compute in one regularization scheme the ratio of say, the leading terms, and in a different regularization scheme the ratio of subleading terms (at same order), and find a different ratio in the two schemes, universality does not hold.  We show an example below.

Let us consider free massless fermions in $d=3$, a conformal field theory. 
For the leading terms, we first perform a numerical computation using similar methods as above, specified in \cite{Casini:2009sr}.%
 The numerical computation leads to
\beq
\frac{C_E}{S_{EE}}\approx 2.9 \ ,
\eeq
which implies that the ratio of the leading terms is approximately $2.9$.

Next we consider subleading terms, and a different regularization scheme. 
For $d=3$, the expansion of R\'enyi entropies contains a universal constant term, related to topological properties of the system. For entanglement entropy, this is known as the $F$-term (or the negation of it). The $F$-term has been shown to be equal to the constant term in the free energy of the system restricted to a unit sphere, $S^3$. This has been computed for free massless fermions in e.g. \cite{Klebanov:2011uf}. The corresponding term for the heat capacity has been computed e.g. in \cite{Perlmutter:2013gua}. The ratio of these two terms is approximately $1.4$, in disagreement with the ratio of leading terms.

Therefore, we can conclude that at least for generic conformal field theories, the ratio of coefficients in the expansions of entanglement entropy and capacity of entanglement are not in general universal.\footnote{There is an exception: the universal coefficients in the expansions are scheme-independent, so there ratios should also be.} 

Narrowing the set of theories further, we could consider CFTs which have a gravity dual. In these cases, geometry may provide a natural regularization scheme. Indeed,
in sections \ref{Highdalt} and \ref{Holosec} we saw that both the entanglement entropy and the capacity of entanglement were proportional to the hyperbolic volume factor containing the 
divergences, leading to a clean ratio of the two. Regarding the interesting question which CFTs then have gravitational duals, we are lead to speculate that perhaps such natural
ratios in CFTs are a hint of a dual gravitational interpretation.

\subsection{On the shape dependence of the capacity of entanglement}

Much of the discussion has focused on spherical or planar entangling surfaces. As a natural generalization, we briefly comment on the shape dependence of the capacity of entanglement.

For general shapes in $d>2$ quantum field theories, the coefficient of the leading divergent term of the R\'enyi entropies is proportional to the area of the entangling surface and non-universal. The subleading terms have more variety and, in general, the expansion is different from the spherical case. Much work has been carried out to understand the shape dependence of entanglement R\'enyi entropies, and some universal results are known. 

A well-studied case is the $4$-dimensional CFTs, for which the universal log term can be expressed in terms of integrals over the entangling surface $\Sigma$, \cite{Fursaev:2012mp}
\beq
S_\alpha^{\rm{univ}}=-\left( \frac{f_a(\alpha)}{2\pi}\mathcal{R}_{\Sigma}+\frac{f_b(\alpha)}{2\pi}\mathcal{K}_{\Sigma}-\frac{f_c(\alpha)}{2\pi}\mathcal{C}_{\Sigma}  \right)\log \left(\frac{1}{\epsilon}\right),
\eeq
where
\bear
\mathcal{R}_{\Sigma}=\int_{\Sigma}d^{2}y\sqrt{h}R_{\Sigma},\quad \mathcal{C}_{\Sigma}=\int_{\Sigma}d^{2}y\sqrt{h}C^{ab}_{\;\;\;ab}, \quad\mathcal{K}_{\Sigma}=\int_{\Sigma}d^{2}y\sqrt{h}\left[\tr K^2-\frac{1}{2}(\tr K)^2   \right].
\eear
Here $y$ and $h$ are the coordinates of the entangling surface and the induced metric, respectively, and $R_{\Sigma}$, $K$, and, $C^{ab}_{ab}$, the intrinsic Ricci scalar of $\Sigma$, the extrinsic curvature of $\Sigma$, and the contraction of the Weyl tensor along coordinates orthogonal to the entangling surface. The functions $f$ depend only on the physical data of the CFT and $n$. The function $f_a$ can be computed by mapping the sphere to hyperbolic spacetime. On the other hand, $f_c$ and $f_b$ can be studied using deformations of spherical entangling surfaces to first and second order, respectively \cite{Dong:2016wcf}. In general, it is known that $f_a(1)=a$, $f_b(1)=f_c(1)=c$. The universal log term of the entanglement
entropy is thus
\beq
S_{EE}^{\rm{univ}}=-\frac{c}{2\pi}\left( \frac{a}{c} \mathcal{R}_{\Sigma}+\mathcal{K}_{\Sigma}-\mathcal{C}_{\Sigma}  \right)\log \left(\frac{1}{\epsilon}\right)  \ .
\eeq
In comparison, the universal log term of the capacity of entanglement for holographic theories is \cite{Dong:2016wcf}
\bear
C_E^{univ}&=&2\left( -\frac{c}{4\pi}\mathcal{R}_{\Sigma}+\frac{f_b'(1)}{2\pi}\mathcal{K}_{\Sigma}-\frac{f'_c(1)}{2\pi}\mathcal{C}_{\Sigma} \right)\log\left(\frac{1}{\epsilon}\right)\\
 &=& -\frac{c}{2\pi}\left( \mathcal{R}_{\Sigma} +\frac{11}{6}\mathcal{K}_{\Sigma}-\frac{17}{9}\mathcal{C}_{\Sigma} \right)\log\left(\frac{1}{\epsilon}\right)
\eear
The ratio $C_E^{univ}/S_{EE}^{\rm{univ}}$ thus depends on $a,c$ and the geometric quantities. 

Similar expressions with integrals over local geometric quantities for the universal terms should exist for other even-dimensional CFTs \cite{Lewkowycz:2014jia}. For odd dimensions, this is not possible for the universal constant term, making the computation of the universal term more challenging.\footnote{For general dimensions, 
there has been interest in \emph{e.g.} the universal term of entanglement entropy by corners and conical singularities \cite{Bueno:2015rda,Bueno:2015qya,Bueno:2015lza} and small deformations \cite{Bianchi:2016xvf}. Interestingly, with the inclusion of conical singularities in the entanglement surface, an additional log term emerges, modifying the form of universal terms for all R\'enyi entropies \cite{Bueno:2015lza}
\beq
S_\alpha^{\rm univ} = \begin{cases}(-1)^{\frac{d-1}{2}}a_\alpha^{(d)}(\Omega)\log(R/\delta) & d\,\,\, {\rm odd}\\
(-1)^{\frac{d-2}{2}}a_\alpha^{(d)}(\Omega)\log^2(R/\delta) & d\,\,\, {\rm even}\\
\end{cases}
\eeq
where the positive $a_\alpha^{(d)}$ depend on the opening angle, $0\leq\Omega\leq\pi$. These kinds of terms would then also appear in the capacity of entanglement.}

For spherical entangling surfaces of CFTs with gravity duals, we saw that there was a natural choice of UV regulation such that the entanglement entropy and the capacity of entanglement were proportional to each other. It would be interesting to see, whether deformed shapes would also have similarly natural choices for UV regulation.

\section{The capacity of entanglement under perturbation with relevant operators}
\label{Sec6}

In previous sections we studied conformal field theories, and in particular in 1+1 dimensions we found that the leading terms of the first cumulants, the capacity of entanglement $\bra K^2\ket_c=C_E$ and the entanglement entropy $\bra K\ket_c=S_{EE}$ were equal. In this section we will study what happens when the theory develops a mass gap or more generally is deformed by
relevant operators to break the conformal invariance.

As a warm-up example, we study the anisotropic Heisenberg XY model. We find that the relationship $C_E = S_{EE}$ is broken, when the 
parameters are moved away from the critical domains, and the equivalent free fermion
system develops a mass gap. We will also study how the divergence structure of the capacity of entanglement alters under perturbing away from criticality.
The first order perturbation will contain a new universal $\log^2$ divergent term. 

After this concrete example, will now perform a more general analysis and study CFTs perturbed with relevant operators. We will be considering two different entangling surfaces, planar and spherical. The strategy that we use follows the work \cite{Rosenhaus:2014zza,Rosenhaus:2014ula}. 

\subsection{The capacity of entanglement in the anisotropic Heisenberg XY spin chain}
\label{XYsec}

The anisotropic Heisenberg XY spin chain is a nice example of a system with a nontrivial phase structure. 
The system can be mapped to noninteracting fermions, with critical domains having a gapless
spectrum, but elsewhere the spectrum is gapped. Using analytical results by Korepin et al \cite{Franchini:2007eu,Franchini:2010kq} on R\'enyi entropy in the anisotropic Heisenberg spin chain,
we compare $S_{EE}$ with the capacity of entanglement. We find that at criticality  $C_E = S_{EE}$, but moving away from the critical domains causes deviations
from this relation.  

The Hamiltonian of the model is
\beq
  H = -\sum^\infty_{^j=-\infty} \left[ (1+\gamma )\sigma^x_j \sigma^x_{j+1} +  (1-\gamma )\sigma^y_j \sigma^y_{j+1} + h\sigma^z_j \right] ,
\eeq
where $\gamma$ parameterizes anisotropy, and $h$ is the external magnetic field. The phase diagram of the model has three regions:
\begin{eqnarray*}
&& {\rm 1a:} \ \ \ \ 4(1-\gamma^2) < h^2 < 4 \\
&& {\rm 1b:} \ \ \ \ h^2 < 4(1-\gamma^2 ) \\
&& {\rm 2:} \ \ \ \ h> 2
\end{eqnarray*}
with the critical lines $\gamma=0, h\leq 2$ where it becomes isotropic (the XX-model), and $h=h_c=2$ corresponding to the critical value of the magnetic field.  The line $h=h_f(\gamma) = 2\sqrt{1-\gamma^2}$ separating the regions $1a,b$ is not a phase transition, but the entanglement entropy has a weak singularity on it,
and ground state is double degenerate with a basis given by two product states.

 Korepin et al computed the entanglement and R\'enyi entropies for a spin chain of $N$ spins, considering the system in its ground state, 
separating a subsystem $A$ of $L$ spins, and considered the double scaling limit $N,L\rightarrow \infty$ with $L/N$ fixed. For the R\'enyi entropy (with the reduced density
matrix $\rho_A$ and exponent $\alpha$) they obtained
\beq\label{Renyixy}
S_R(\rho_A,\alpha ) = \left\{ \begin{array}{l} \frac{1}{6}\frac{\alpha}{1-\alpha} \ln (k~k') -\frac{1}{3}\frac{1}{1-\alpha} 
\ln \left( \frac{\theta_2(0,q^\alpha )\theta_4 (0,q^\alpha )}{\theta^2_3 (0,q^\alpha )}\right) +\frac{1}{3} \ln 2 \ , \ h>2 \\
 \frac{1}{6}\frac{\alpha}{1-\alpha} \ln (\frac{k'}{k^2}) -\frac{1}{3}\frac{1}{1-\alpha} 
\ln \left( \frac{\theta^2_3 (0,q^\alpha ) }{ \theta_2(0,q^\alpha )\theta_4 (0,q^\alpha )}\right) +\frac{1}{3} \ln 2 \ , \ h<2 \end{array} \right.
\eeq
with the elliptic modulus parameter
\beq
 k = \left\{ \begin{array}{l} \sqrt{(h/2)^2+\gamma^2-1} / \gamma \ \ \ \ {\rm :1a} \\
\sqrt{(1-h^2/4-\gamma^2)/(1-h^2/4)} \ \ \ \ {\rm :1b} \\
\gamma / \sqrt{(h/2)^2+\gamma^2-1} \ \ \ \ {\rm :2}
\end{array}  \right.
\eeq
and its complement $k' = \sqrt{1-k^2}$, and the nome
\beq
 q = e^{-\pi I(k')/I(k)} \equiv e^{-\pi \tau_0}
\eeq
where $I(k)$ is the complete elliptic integral of the first kind. Either by taking the limit $\alpha =1$, or computing the first derivative of the generating function
$\tilde{k}(\alpha ) \equiv (1-\alpha)S_R(\rho_A,\alpha)$, the entanglement entropy becomes
 \beq
S_{EE}(\rho_A) = \left\{ \begin{array}{l} \frac{1}{6} \left[\ln \frac{4}{k~k'} + (k^2-k'^2)\frac{2I(k)I(k')}{\pi})\right] \  , \ \ \ h> 2 \\
\frac{1}{6} \left[\ln \left( \frac{4k^2}{k'}\right) + (2-k^2)\frac{2I(k)I(k')}{\pi})\right] \  , \ \ \ h< 2 \end{array} \right.
\eeq
 The result for $h>2$ comes from an expression
\beq
  S_{EE} (\rho_A) = \frac{1}{6} \ln\frac{4}{k~k'} +\frac{1}{12} \ln q + 2\ln q \sum^\infty_{m=0} \frac{(2m+1)q^{2m+1}}{1+q^{2m+1}} \ ,
\eeq
where the latter series can be found from Abramowitz and Stegun in closed form with the elliptic integrals \cite{Peschel} to recover the result in the above.
In region $h>2$, the capacity of entanglement, from the double derivative of the generating function, becomes
\beq
C_{E} (\rho_A) = 2\ln q \sum^\infty_{m=0} \left[\frac{(2m+1)q^{2m+1}}{1+q^{2m+1}}\right]^2 \ ,
\eeq
We have not identified the closed form for the infinite series, but appears clear that in general the capacity of entanglement differs from $S_{EE}$.  We can study this in more 
detail near the critical phases, to find that at criticality $C_{E} = S_{EE}$ while they begin to deviate moving away from the critical phase. 

For $\gamma \neq 0$, near the phase boundary at $h_c=2$, for the leading and next-to-leading order contributions we obtain
\bear\label{hc2}
S_{EE}(\gamma,h) &=& \frac{1}{6}\log\left(\frac{16\gamma^2}{|h-2|}\right)+\left(\frac{5-\gamma^2-3\log\left(\frac{16\gamma^2}{|h-2|}\right)}{24\gamma^2}\right)(h-2)
\nonumber \\ &&+ \mathcal{O}((h-2)^2)\\
C_E (\gamma,h) &=&  \frac{1}{6}\log\left(\frac{16\gamma^2}{|h-2|}\right)+\left(\frac{2-\gamma^2-6\log\left(\frac{16\gamma^2}{|h-2|}\right)+3\log\left(\frac{16\gamma^2}{|h-2|}\right)^2}{24\gamma^2}\right)(h-2) \nonumber \\
&&+ \mathcal{O}((h-2)^2) \label{eq:pert-magnetxy} \ ,
\eear
where the expansion parameter is the inverse of the relevant length scale $\xi$,
\beq
    \xi^{-1} = |h-2| \ .
\eeq
The leading term in the entanglement entropy matches with the result for a conformal field theory with $c=1/2$,
\beq\label{scftxi}
  S_{EE} = \frac{c}{3}\log \xi \ .
\eeq
 We can now investigate how the deviation from  $C_{E} = S_{EE}$  happens in the vicinity of the critical line. 
The difference of the two, $S_{EE}-C_{E}$ is depicted in the Figure 3A. In this region $S_{EE}$ grows faster than the capacity of entanglement $C_{E}$.

\begin{figure}[h]
\begin{center}
\includegraphics[width=0.4 \textwidth]{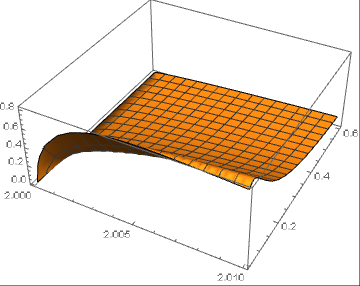}
\hfil
\includegraphics[width=0.4 \textwidth]{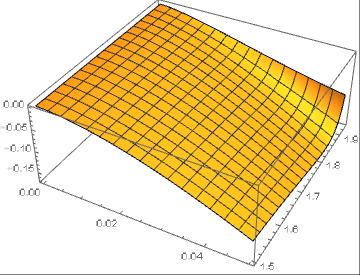}
\\
({\bf A})
\hfil ({\bf B})
 \\
\caption{\label{fig:ssvarxy}
Samples of the difference $S_{EE}-C_{E}$ near the critical line $h_c=2$ ({\bf A}) and near the critical line $\gamma =0$ ({\bf B}) .}
\label{SdeltaS}
\end{center}
\end{figure}

On the other hand, near the isotropic critical line $\gamma=0$, for $h<2$, we find
\bear
S_{EE}(\rho_A) &=& \frac{1}{6} \log\left(\frac{4 \left(4-h^2\right)}{\gamma ^2}\right)
+ \frac{\left(5+3 \log \left(\frac{\gamma ^2}{4 \left(4-h^2\right)}\right)\right)}{6 \left(h^2-4\right)}\gamma ^2\nonumber \\&&\quad+\mathcal{O}(\gamma^4) \\
C_{E}(\rho_A) &=& \frac{1}{6} \log\left(\frac{4 \left(4-h^2\right)}{\gamma ^2}\right)-\frac{\left(2+3 \log ^2\left(\frac{\gamma ^2}{16-4 h^2}\right)+6 \log \left(\frac{\gamma ^2}{16-4 h^2}\right)\right)}{6\left(4-h^2\right)}\gamma^2 \nonumber\\
&& \quad+\mathcal{O}(\gamma^4) \ ,\label{eq:pert-isotropicxy}
\eear
with the relevant inverse length scale
\beq
 \xi^{-1} = |\gamma | \ ,
\eeq
so the leading term in $S_{EE}$ matches with the CFT result (\ref{scftxi}) with $c=1$.
Figure 3B depicts $S_{EE}-C_{E}$ in this case. In this region $C_{E}$ grows faster than $S_{EE}$.  The saddle seen
in figure \ref{fig:ssvarxy} likely reflects the boundary $h_f(\gamma)=2\sqrt{1-\gamma^2}$ separating the phase regions 1a and 1b, but a full matching would require a higher order calculation
than (\ref{hc2}). From the approximate results it appears that $S_{EE}$ is larger in the vicinity of the $h_c=2$ critical line, while $C_{E}$ is larger in the vicinity
of the $\gamma =0$ critical line.

\subsection{Field theories perturbed with relevant operators}

Much of the initial discussion here has already appeared in \cite{Rosenhaus:2014woa}. Let $\mathcal{M}$ be a $d$-dimensional Euclidean manifold. Consider its ground state $|0\rangle$ and the corresponding density matrix $|0\rangle\langle 0|$. We partition the manifold to $V$ and ${\bar V}$ and take a trace of the density matrix over the degrees of freedom in ${\bar V}$, obtaining the reduced density matrix in $V$, $\rho_V$. Inspired by the replica trick, we consider a cut $\mathcal{C}$ and its both sides, $\mathcal{C}_+$ and $\mathcal{C}_-$. The density matrix $\rho$ can be expressed as a path integral
\beq
[\rho_V]_{\phi_-\phi_+} = \frac{1}{\mathcal{N}}\int_{\footnotesize{\begin{array}{c}\phi(\mathcal{C}_-)=\phi_-\\\phi(\mathcal{C}_+)=\phi_+\end{array}}}{\rm e}^{-I[\phi]}.
\eeq

Now, we let $I$ be an action with a small perturbation by operator $O$, i.e. $I[\phi]=I_0[\phi]-g\int_{\mathcal{M}} O$.  Note that the action governs the whole system, so the integral is over the whole manifold $\mathcal{M}$, not just $V$. In the limit of small $g$, we expand
\beq
[\rho_V]_{\phi_-\phi_+} = \frac{1}{\mathcal{N}}\int_{\footnotesize{\begin{array}{c}\phi(\mathcal{C}_-)=\phi_-\\\phi(\mathcal{C}_+)=\phi_+\end{array}}}{\rm e}^{-I_0[\phi]}\left(1+g\int_{M}O+\frac{g^2}{2}\int_{\mathcal{M}}\int_{\mathcal{M}}OO+\ldots\right)\,,
\eeq
with
\beq
\mathcal{N}=\mathcal{N}_0\left(1+ g\int_{\mathcal{M}} \langle O\rangle_0+\frac{g^2}{2}\int_{\mathcal{M}}\int_{\mathcal{M}}\langle OO\rangle_0+\ldots\right) \,,
\eeq
where $\langle \cdots \rangle_0$ is the vacuum expectation value in the unperturbed theory. To first order, 
\bear
[\rho_V]_{\phi_-\phi_+}&\approx&\frac{1}{\mathcal{N}_0}\int_{\footnotesize{\begin{array}{c}\phi(\mathcal{C}_-)=\phi_-\\\phi(\mathcal{C}_+)=\phi_+\end{array}}}{\rm e}^{-I_0[\phi]}(1+g\int_{\mathcal{M}}(O-\langle O\rangle_0 ))\\
 &=& \left[\rho_{V,0}+g \int_{\mathcal{M}}\Tr_{{\bar V}}\rho_0(O-\langle O\rangle_0 )\right]_{\phi_-\phi_+} \,,
\eear
where $\rho_{0}$ and $\rho_{V,0}$ are the unperturbed vacuum density matrices of the whole system and the subsystem, respectively. The vacuum expectation value of $O$ vanishes whenever $O$ is a pseudoprimary operator. We assume this to be the case from now on. In addition, we focus on operators with scaling dimension $\Delta>\frac{d}{2}$ as the perturbative expansion of CFTs fail whenever $\Delta\leq \frac{d}{2}$ \cite{Nishioka:2014kpa}.

To compute the R\'enyi entropies, we need
\beq
\Tr \rho_V^n = \Tr \rho_{V,0}^n+ng\int_{\mathcal{M}}\Tr \left[\rho_{V,0}^{n-1}\Tr_{\bar V}\rho_0 O\right] + \mathcal{O}(g^2) \ .
\eeq
The first order correction to the entanglement entropy is then
\beq
-g\partial_g\partial_n \Tr \rho_V^n|_{n= 1,g= 0}=g\int_{\mathcal{M}} \Tr \left[K \Tr_{\bar V}\rho_0 O\right] \,,
\eeq
and to the capacity of entanglement,
\beq
g\partial_g\partial_n^2 \log\Tr \rho_V^n|_{n= 1,g= 0}=g\int_{\mathcal{M}} \Tr \left[(K^2-2K) \Tr_{\bar V}\rho_0 O\right]-2\Tr\left[K\right]\Tr\left[K\Tr_{\bar V}\rho_0 O\right].
\eeq
In the case of a QFT with a planar entangling surface or a CFT with a spherical entangling surface, $K$ is known and expressible as an integral of the energy-momentum tensor. As a final remark, the last term in the perturbation of variance can be ignored as its contributions will cancel with the normalization constants of the former terms.

\subsection{Planar entangling surface}
We consider first a planar entangling surface $\Sigma = \mathbb{R}^{d-2}$, at $x_1=0, x_0=0$. For any Cauchy surface $A$ that ends at the entangling surface, the modular Hamiltonian is\footnote{For the sign, we follow the conventions of \cite{Rosenhaus:2014woa}.}
\beq
K = -2\pi\int_A {\rm d}^{d-2}y \,{\rm d} x_1n^{\mu} T_{\mu\nu}\xi^{\nu},
\eeq
where $x_0$ and $x_1$ are the coordinates transverse to the entangling surface and $y$ are the coordinates parallel to $\Sigma$. In addition, $\xi = x_1\partial_{x_0}-x_0\partial_{x_1}$ is a Killing vector that keeps the entangling surface invariant and $n$ is the normal vector to the Cauchy surface.

The usual choice for the Cauchy surface is to set $x_0=0$ which yields
\beq
K = -2\pi \int_{\Sigma} d^{d-2} y\int\limits_{0}^{\infty} dx_1 x_1 T_{00},
\eeq
Due to the freedom in choosing the Cauchy surface, we can also write $K$ as
\beq
K = 2\pi \int_{\Sigma} d^{d-2} y\int\limits_{-\infty}^{0} dx_1 x_1 T_{00}
\eeq
In addition, we can also rewrite the integral of the $O$ contribution in the $\langle K O\rangle$ term as \cite{Rosenhaus:2014ula}
\beq
\int_{\mathbb{R}^d}{\rm d}^{d}x\,O\to 2\pi \int_{\Sigma} d^{d-2} y\int\limits_{0}^{\infty} dx_1 x_1\, O.
\eeq

\subsubsection{Massive free scalar field theory in four Euclidean dimensions}

As in \cite{Rosenhaus:2014zza}, we start with the free scalar field theory, with a mass term as a relevant deformation: 
$g=-\frac{m^2}{2}$ and $\mathcal{O}(x)= \phi^2(x)$. The stress tensor of the massless theory is
\beq
T_{\mu\nu}^0 =\partial_{\mu}\phi \partial_{\nu}\phi-\frac{1}{2}\delta_{\mu\nu}(\partial \phi)^2.
\eeq
Using Wick contractions, we can compute the terms appearing in the variation of variance. The two-point functions of the scalar fields is
\beq
\langle\phi(x)\phi(y)\rangle = \frac{1}{(d-2)\Omega_{d-1}}\frac{1}{(x-y)^{d-2}}.
\eeq
We point the reader to the appendices for computational details.

When computing the integrals, we will run into UV or IR divergences. These emerge when integrating perpendicular to the entangling surface. We will regulate these with $\varepsilon$ and $1/m$ respectively. Summing all the contributions, we get
\bear
\delta C_{E} &=&g\mathcal{A}_{\Sigma}\frac{ (d-3)  \pi ^{\frac{3-d}{2}} \Gamma \left(\frac{d}{2}-1\right)^2}{2^{d-1}(d-2)(d-4) \Gamma \left(\frac{d+1}{2}\right)}\left(\delta^{4-d}-m^{d-4}\right)\\
&\stackrel{d=4}{=}&-g\frac{\mathcal{A}_{\Sigma}}{12 \pi}\log(\delta m)
\eear
so in $d=4$ the first order correction to the capacity of entanglement $\Delta S^2_{EE}$ is
\beq
\delta C_{E} \stackrel{d=4}{=} \frac{m^2}{24\pi}\mathcal{A}_{\Sigma}\log(m\delta).
\eeq
In the above, $\mathcal{A}_{\Sigma}$ is the area of the entangling surface.

Interestingly, the first order perturbation to entanglement entropy has the form
\bear
\delta S_{EE} &=& g\mathcal{A}_{\Sigma}\frac{\pi^{\frac{3-d}{2}}\Gamma\left(\frac{d}{2}\right)^2}{2^{d-1}(d-2)(d-4)\Gamma\left(\frac{d+1}{2}\right)}\left(\delta^{4-d}-m^{d-4}\right) \\
&\stackrel{d= 4}{=}&-\frac{g\mathcal{A}_{\Sigma}}{12 \pi}\log(m\delta).
\eear
Thus, when $d=4$, the correction terms are equal for entanglement entropy and the capacity of entanglement.

\subsubsection{A general CFT}\label{subsec:gencft}
The computation for a general CFT is more involved. We need the general forms of $\langle T_{\mu\nu} \Ocal \rangle$ and $\langle T_{\mu\nu} T_{\alpha\beta}\Ocal \rangle$ for any CFT. Due to the tracelessness and sourcelessness of the energy-momentum tensor, the two-point function automatically vanishes for any CFT. Thus, the entanglement entropy
receives no correction at first order in the coupling $g$. This is not the case for the capacity of entanglement. The computation requires some effort but is, nevertheless, straightforward. We leave the computational details to appendices and move on to the results.

We consider two special cases. In the following, $a_3$ is a multiplicative numerical factor appearing in the three point function. In most cases, we do not know any physical interpretation for it.

First, for $\Delta = 2$ and all $d$ we get a logarithmic contribution. In this case, the general first order correction to the capacity of entanglement is %
\beq
\delta C_{E} = ga_3 \frac{4 (d-2) \pi ^{d+1} \cos \left(\frac{\pi  d}{2}\right) \Gamma (3-d) \log (\delta  \, \Lambda)}{d(d-2) (d-3) -4}\mathcal{A}_{\Sigma},
\eeq
where $\Lambda$ is an IR regulator and $a_3$ is a numerical coefficient. Interestingly, the result has an IR divergence only for even $d\geq 4$, in agreement with the breakdown of CFT perturbations for $\Delta\leq\frac{d}{2}$.  For $d=2$, the correction
vanishes. In fact, $\bra T_{\mu\nu}T_{\alpha \beta}O \ket =0$ at $d=2$, so just like $S_{EE}$, $C_E$ receives no corrections at leading order, for all $\Delta$ for which
the perturbative approach is valid.

Second, we consider the weight $\Delta=d-2$ which corresponds {\em e.g.} to a mass term in a scalar field theory. The first order correction is %
\beq\label{etad2}
\delta C_{E}= ga_3 \frac{32 (d-3) \pi ^{d+1}}{d(d-2)(d-4)^2 \Gamma (d-1)}\left(\Lambda^{d-4}-\delta^{4-d}\right)\mathcal{A}_{\Sigma}.
\eeq
Once again, we see the emergence of the divergences near $d=4$. However, the divergence for $d=2$ is not a real divergence as the three-point function is zero. The expression vanishes when $d=3$.

The overall numerical factor $a_3$, for a conformally coupled scalar field theory with a mass perturbation $g = -\frac{m^2}{2}$,  has the value
\beq
a_3 = \frac{d}{2(d-1)^2\Omega_{d-1}^3}=\frac{d[\Gamma\left(\frac{d}{2}\right)]^3}{16\pi^{\frac{3d}{2}}(d-1)^2}
\eeq

To summarize, in these special cases we find divergent corrections $\delta C_{E}$ at leading order in $g$. The exceptions are $d=3, \Delta =1$ and $d=2, \Delta > 1$, where
$\delta C_E =0$.  In particular the equality $C_E=S_{EE}$ receives no corrections at leading order in $g$.

\subsection{Spherical entangling surfaces of a CFT}
There is a conformal mapping from the Rindler wedge to the causal diamond of a ball. Hence, for a spherical entangling surface of radius $R$ in Euclidean spacetime, the modular Hamiltonian for a CFT is
\beq
K = -2\pi\int_{B'} n^{\mu} T_{\mu\nu}\xi^{\nu},
\eeq
where $B'$ is any Cauchy surface that ends at the spherical entangling surface. Once again, $n$ is the normal vector to the Cauchy surface and $\xi=\frac{1}{2R}(R^2+x_0^2-r^2)\partial_0+\frac{x_0 r}{R}\partial_r$ is a conformal Killing vector that keeps the entangling surface at $x_0=0$, $r= R$ invariant. With the simple choice $B'=B$, the modular Hamiltonian is
\beq
K = -2\pi \int_{B(0,R)} {\rm d}^{d-1}x \frac{R^2-r^2}{2R}T_{00}(0,x).
\eeq

When computing the effects of the pertubation, only the $\langle KKO\rangle$ term is found to be non-zero at first order. We can recycle some of the computations done for the planar entangling surface. We leave many of the details to the appendices. The correction to the capacity of entanglement is
\bea
\delta C_E\!\!\! &=&\!\!\! g\int_{\mathbb{R}^d}{\rm d}^d x_3\langle KKO(x_3)\rangle\\
\!\!\!\!\!&=&\!\!\!g(2\pi)^2 \int_{|x_1|\leq R}\!\!\!\!\! {\rm d}^{d-1}x_1\int_{|x_2|\leq R}\!\!\!\!\! {\rm d}^{d-1}x_2\int_{\mathbb{R}^d}\!\!\!{\rm d}^dx_3\frac{(R^2-r_1^2)(R^2-r_2^2)}{4R^2}\langle T_{tt}(0,x_1)T_{tt}(0,x_2) O(x_3)\rangle\nonumber.
\eea
The integral over the $x_3$ coordinates is essentially the same as for the planar entangling surface done in \ref{gencft}.

After the integral over $x_3$, we are left with
\beq\label{leftwith}
\delta C_E = ga_3(2\pi)^2\pi^{d/2}f(d,\Delta)\frac{\Gamma\left(\frac{d-\Delta}{2}\right)^2\Gamma\left(\Delta-\frac{d}{2}\right)}{\Gamma\left(\frac{\Delta}{2}+2\right)\Gamma\left(d-\Delta\right)}~I_d %
\eeq
where $a_3$ is the numerical factor that already appeared in the planar computations and $f$ is a rational function and its specific value for arbitrary $d$ and $\Delta$ is seen in \eqref{eq:ffunc}. The gamma functions capture the possible IR divergences for $\Delta\leq d/2$ (where the perturbative approach breaks down \cite{Nishioka:2014kpa}). In addition, the relevant perturbations are for $\Delta \leq d$.
What remains is the computation of the integral $I_d$,
\beq
I_d = \int_{|x_1|\leq R}\!\!\!\!\! {\rm d}^{d-1}x_1\int_{|x_2|\leq R}\!\!\!\!\! {\rm d}^{d-1}x_2\frac{(R^2-r_1^2)(R^2-r_2^2)}{4R^2}\frac{1}{(x_1-x_2)^{d+\Delta}} \ . \label{eq:xintegral}
\eeq
Further details on evaluating this integral can be found in appendix \ref{sphereintegral}. 

The perturbation has a universal logarithmic term whenever $\Delta$ is an even number. It is
\beq
\delta C^{(log)}_E=-ga_3f(d,\Delta)\frac{\pi ^{\frac{3 d}{2}-\frac{1}{2}} 2^{2-\Delta } \Gamma \left(\frac{-\Delta -1}{2} \right) \Gamma \left(2 \Delta -\frac{1}{2}\right) \Gamma\left(\frac{d-\Delta }{2}\right)^2 \Gamma \left(\Delta -\frac{d}{2}\right) R^{d-\Delta }}{\Gamma \left(\frac{d-1}{2}\right) \Gamma \left(\frac{\Delta }{2}+2\right)^2 \Gamma (2 \Delta ) \Gamma (d-\Delta ) \Gamma \left(\frac{1}{2} (d-\Delta +4)\right)}\log(\epsilon).
\eeq
As a specific example, consider $d=4$, then
\beq
\delta C^{(log)}_{E,d=4}= ga_3\frac{2 \pi ^{11/2} \Delta (\Delta+2) R^{4-\Delta} \Gamma (\Delta-3) \Gamma \left(2\Delta-\frac{1}{2}\right)}{(3 (\frac{\Delta}{2}-4)\frac{\Delta}{2}+10) \Gamma (2 \Delta) \Gamma (\frac{\Delta}{2}+2)^2}\log(\epsilon).
\eeq

For other values of $\Delta$, there is a universal constant term, which is
\beq
\delta C^{(const)}_E= ga_3\frac{f(d,\Delta)2^{8-d} \pi ^{\frac{3 d}{2}+1} \Gamma \left(1-\frac{\Delta }{2}\right) \Gamma \left(\Delta -\frac{d}{2}\right) R^{d-\Delta } }{\left(\Delta ^2-1\right)(d-\Delta)(d-\Delta+2)\Gamma \left(\frac{d-1}{2}\right)\Gamma\left(\frac{\Delta }{2}+2\right)^2\Gamma \left(\frac{1}{2} (d-\Delta +1)\right)} \ .
\eeq
Once again, we consider an example, $d=4$, for which
\beq
\delta C^{(const)}_{E,d=4}=ga_3\frac{8 \pi ^{13/2} \Delta  (\Delta +2) \Gamma \left(1-\frac{\Delta }{2}\right) \Gamma (\Delta -2) R^{4-\Delta }}{(3 (\Delta -8) \Delta +40) \Gamma\left(\frac{5}{2}-\frac{\Delta }{2}\right) \Gamma \left(\frac{\Delta }{2}+2\right)^2} \ .
\eeq

To summarize, we have isolated explicit expressions for universal perturbative corrections to $C_E$ in $d=4$. Similar results can be obtained in other dimensions from (\ref{leftwith}).

\section{Discussion and outlook}
\label{Sec8}

In this paper we discussed several aspects of capacity of entanglement $C_E$, which encodes a particular feature
of the set of eigenvalues of a reduced density matrix. It can roughly be thought of as the variance of
the eigenvalue distribution, and can be obtained in a straightforward way from the (analytically continued)
R\'enyi entropies. It may have interesting divergences across quantum phase transitions (in which case particular
critical exponents may appear) but we have not studied this particular aspect in this paper.

If the reduced density matrix is viewed as a thermal density matrix, and we are allowed to change its
`temperature' by considering a suitable one-parameter flow (the "modular flow" $\rho (\theta ) = \rho^{1+\theta}
 / \Tr (\rho^{1+\theta})$), then we can relate capacity of entanglement to the ordinary heat capacity defined
 with respect to this fiducial temperature. At the same time this allows us to connect to various
 information theoretic quantities like fidelity susceptibility and Fisher information. Unfortunately,
in an actual generic physical subsystem there is no natural operation corresponding to this change in
`temperature' and therefore these relations remain somewhat academic.

While in general $C_E$ can by much larger than $S_{EE}$, we found several situations where $C_E$ is comparable
to $S_{EE}$. First and foremost this happens in CFT's with a holographic dual where both $S_{EE}$ and $C_E$
have an area law divergence. In principle the coefficients that appear in this divergence are scheme dependent
in $d>2$, but if the reduced density matrix is that associated to a ball in the ground state there is a fairly
natural scheme in which $S_{EE}$ is precisely equal to $C_E$. 

We have studied various setups in order to gain some intuition for this approximate equality. First,
following earlier work, we related the capacity of entanglement to fluctuations of $U(1)$ charges fluctuating in
and out of the subregion, communicating entanglement. In this context, if the fluctuations are Gaussian distributed, 
or if the total particle number $N\rightarrow \infty$, we found that the capacity of entanglement becomes 
equal to the entanglement entropy, $C_{E}= S_{EE}$.

In a second setup, which we expect to be closely related to the previous one with fluctuation $U(1)$ charges,
we considered random bipartite entanglement and observed that for randomly entangled pairs of qubits $C_E$ 
and $S_{EE}$ are also approximately equal.

All this then suggests that the approximate equality of $C_E$ and $S_{EE}$ implies that entanglement is
effectively carried by randomly entangled UV pairs of qubits and not by large numbers of maximally entangled
EPR pairs, as in the latter case $C_E$ would be much smaller than $S_{EE}$. This in particular applies
to the quasiparticle picture sometimes used to describe entanglement growth after quenches: if $C_E$ is
proportional to $S_{EE}$, the quasiparticles should be approximately randomly entangled and not in
maximally entangled EPR pairs. 

The area law for capacity of entanglement seems to extend quite generally to an area law for R\'enyi entropies.
This is perhaps not that surprising, because a volume law would disagree with the basic observation that
the R\'enyi entropy for a subregion equals that of the complement of the subregion. In addition, 
such an area law is also in agreement with the above picture of randomly entangled UV pairs of qubits, and as
we explained in section~\ref{enspec}, in agreement with a local Rindler picture of the entangled degrees of freedom. 
This is all self-consistent, as in Rindler space the entanglement is thermally distributed and therefore
also not mostly carried by maximally entangled EPR pairs.

In AdS/CFT, $C_E$ has a relatively simply bulk interpretation, given by metric fluctuations integrated
over the Ryu-Takayanagi surface. This relationship is strongly reminiscent of the relation 
between response functions and fluctuations in Landau-Ginzburg theories. It is not entirely obvious
where the equality $C_E \sim S_{EE}$ comes from in this computation. One would expect a divergence
to arise if one or both points in the graviton propagator approach the boundary of the Ryu-Takayanagi
surface. If one point approaches the boundary there is an area divergence from the integral over that
boundary point, but that gets reduced by the graviton propagator which decays at long distances. Therefore
the divergence must come from the region of integration where both points are close to the boundary\footnote{This argument concerns the leading
divergence. The equality of $C_E$ and $S_{EE}$ in quenches (in 2d CFT) indicates that contributions from deeper in the bulk (where the RT surface
crosses the collapsing shell) are also important.}.
Some naive power counting suggests that this gives indeed the right behavior, but it would be nice
to examine this in more detail and try to connect it to the previous discussion.

One of the motivations for this work was to study quantum fluctuations in the metric at the RT surface and
to study the validity of the semiclassical approximation. Since most of the contributions seem to come
from the region near the boundary, capacity as we have defined it is perhaps not a very good probe of
the size of bulk metric fluctuations. A better
probe would be to consider fluctuations in manifestly finite quantities such as mutual information and
relative entropy. It is straightforward to find a generalization of $C_E$ for the case of relative 
entropy\footnote{
While this expression is fairly obvious as it stands, it can also be obtained from various one-parameter
generalizations of the relative entropy called R\'enyi divergences \cite{ren1,ren2,ren3,ren4} 
by differentiating and setting the
parameter equal to 1, similar to what we did for capacity of entanglement. This connection may be 
helpful in order to e.g. establish monotonicity properties of $C_E(\rho|\sigma)$. In CFTs and their gravity
duals,  R\'enyi divergences 
were recently studied in  \cite{Bernamonti:2018vmw}.}. 
It is given by
\begin{equation}
C_E(\rho|\sigma)={\rm Tr}(\rho(\log\rho-\log\sigma)^2) - ({\rm Tr}(\rho(\log\rho-\log\sigma)))^2
\end{equation}
and a candidate quantity that generalize $C_E$ to mutual information is $C_E(\rho_{AB} |
\rho_A\otimes \rho_B)$. It would be interesting to study these quantities in more detail but we
leave that to future work.

The equality of $C_E$ and $S_{EE}$ for (suitably regularized) CFT's with a holographic dual suggests
that it may be possible to extract a necessary criterion for the existence of a holographic dual from 
consideration of capacity of entanglement. This led us to study what happens when one breaks conformal
invariance and we found that the capacity of entanglement of the entanglement entropy starts to deviate from the 
entanglement entropy. As a demonstration, we considered the anisotropic Heisenberg XY spin chain, 
where the equality of $C_E$ and $S_{EE}$ holds at criticality but is broken as the parameters move from the 
critical lines. We also studied deformations of CFTs by relevant operators, and found
similar results. In that case, the analysis is complicated by various singularities whose precise understanding
we leave to future work. 

Finally, we notice that finding estimates of fluctuations in the reduced density matrix as we dial external parameters is
also of great relevance in putting bounds on the accuracy of local measurement. We intend to explore this
connection in more detail in the near future.

\bigskip

\noindent
{\large \bf Acknowledgments}

\medskip

JJ and EKV are in part supported by the Academy of Finland grant no
1297472. JJ is also in part supported by the U. Helsinki Graduate School PAPU, and EKV is also in part supported by a grant from the Wihuri Foundation.  
We thank U. Danielsson, N. Jokela, A. Kupiainen, A. Lawrence, D. Sarkar, W. Taylor, E. Tonni, T. Wiseman for useful discussions and comments while this work was in progress. JdB and EKV also thank the workshops "Quantum Gravity, String Theory, and Holography", "Black Holes and Emergent Spacetime", "Field theory anthropics and naturalness in cosmology", and 
"Holography: What is going on these days?" for hospitality and
partial support during this work.

\setcounter{equation}{0}

\appendix
\section{Computational details of the perturbative computations with planar entangling surface}
\subsection{Mass deformation of a massless free scalar field theory}
First, we need
\beq
\langle\mathcal{O}(x)T^0_{\mu\nu}(z)\rangle = \frac{2(x-z)_{\mu}(x-z)_{\nu}-\delta_{\mu\nu}(x-z)^2 }{\Omega_{d-1}^2(x-z)^{2d}},
\eeq
where $\Omega_{d-1}=\frac{2\pi^{\frac{d}{2}}}{\Gamma(d/2)}$ is the surface area of the $d-1$ sphere and
\bear
&&\langle\mathcal{O}(x)T^0_{\mu\nu}(z)T^0_{\alpha\beta}({\bar z})\rangle = \frac{2}{\Omega_{d-1}^3}\frac{1}{(x-z)^d(x-{\bar z})^d(z-{\bar z})^d} \times\\
&&\Bigg(  (x-z)_{\mu}(x-{\bar z})_{\alpha}\left[\delta_{\nu\beta}-d\frac{ (z-{\bar z})_{\beta} (z-{\bar z})_{\nu}) }{ (z-{\bar z})^2 } \right] + \rm{permutations\, of \,(\alpha,\beta,\mu,\nu)}\nonumber\\
  &&+\delta_{\alpha\beta}\delta_{\mu\nu} \left[ (x-z) \cdot (x-{\bar z}) -d\frac{ ((x-{\bar z}) \cdot (z-{\bar z}) )((x-z) \cdot (z-{\bar z}) )}{(z-\bar{z})^2}     \right] \nonumber\\  
  && -\delta_{\mu\nu} \left[ (x-{\bar z})_{\alpha}(x-z)_{\beta}-d\frac{(x-{\bar z})_{\alpha} (z - \bar{z})_{\beta} [(x-z)\cdot (z-{\bar z}) ] }{(z-{\bar z})^2}  +  (\alpha \leftrightarrow \beta)    \right] \nonumber\\
&&  -\delta_{\alpha\beta}\left[  (x-z)_{\mu}(x-{\bar z})_{\nu}-d\frac{(x-z)_{\mu}(z - \bar{z})_{\nu}[(x-z)\cdot (z-{\bar z})]}{(z-{\bar z})^2} + (\mu \leftrightarrow \nu)   \right] \Bigg).\nonumber
\eear

When we put in $\alpha,\beta,\mu,\nu=0$ and set the stress energy tensor at the same point in time, we get
\bear
\langle\mathcal{O}(x)T^0_{00}(z)T^0_{00}({\bar z})\rangle &=& \frac{2}{\Omega_{d-1}^3}\frac{1}{(x-z)^d(x-{\bar z})^d(z-{\bar z})^d} \label{eq:TTO}\\
&&\times\left[(x-z)\cdot (x-{\bar z})-d\frac{(x-{\bar z})\cdot (z-{\bar z})((x-z)\cdot(z-{\bar z}))}{(z-{\bar z})^2}\right]. \nonumber
\eear

We now compute the required integrals, following \cite{Rosenhaus:2014zza}. We begin with the two-point function as it is simpler. Using time translation symmetry, we set the integral to the form
\bea
\langle \mathcal{O} K \rangle\!\!\!&=&\!\!\! (2\pi)^2 \int_{\Sigma}\!\!\!d^{d-2} y\int_{\Sigma}\!\!\!d^{d-2}{\bar y} \int\limits_{0}^{\infty}dx_1\int\limits_{-\infty}^{0}d{\bar x}_1 \langle O(x) T_{00}({\bar x})\rangle\\
\!\!\!&=&\!\!\! \int_{\Sigma}\!\!\!d^{d-2} y\int_{\Sigma}\!\!\!d^{d-2}{\bar y} \int\limits_{0}^{\infty}dx_1\int\limits_{-\infty}^{0}d{\bar x}_1 \frac{-x_1{\bar x}_1 (2\pi)^2}{\Omega_{d-1}^2((x_1-\bar{x}_1)^2+(y-{\bar y})^2)^{d-1}}.
\eea
We first shift $y\to y+{\bar y}$ and compute the integral over $y$ and $\bar{y}$
\bear
\langle \mathcal{O} K \rangle &=& \frac{-(2\pi)^2\Omega_{d-3}\sqrt{\pi}\Gamma[\frac{d}{2}-1]}{2^{d-1}\Gamma[\frac{d}{2}-\frac{1}{2}]\Omega_{d-1}^2}\mathcal{A}_{\Sigma} \int\limits_{0}^{\infty}dx_1\int\limits_{0}^{\infty}d{\bar x}_1 \frac{x_1{\bar x}_1}{(x_1-{\bar x}_1)^{d}} \\
&=& \frac{(2\pi)^2\Omega_{d-3}\sqrt{\pi}\Gamma[\frac{d}{2}-1]}{(d^2-3d+2)2^{d-1}\Gamma[\frac{d}{2}-\frac{1}{2}]\Omega_{d-1}^2}\mathcal{A}_{\Sigma} \int\limits_{0}^{\infty}dx_1 \frac{1}{x_1^{d-3}}.
\eear

We compute the integral of $x_1$ only from $\delta$ to $m^{-1}$ to regulate both the UV and IR divergences. We take the limit $d\to 4$, yielding
\beq
\mathcal{A}_{\Sigma}\frac{\pi^{\frac{3-d}{2}}\Gamma\left(\frac{d}{2}\right)^2}{2^{d-1}(d-2)(d-4)\Gamma\left(\frac{d+1}{2}\right)}\left(\delta^{4-d}-m^{d-4}\right) \stackrel{d\to 4}{=}\frac{-\mathcal{A}_{\Sigma}}{12 \pi}\log(m\delta).
\eeq

This computation also gives the first order correction to the entanglement entropy $S_{EE}$. In $d=4$ the result is
\beq
\delta S_{EE} = \frac{m^2}{24 \pi}\mathcal{A}_{\Sigma}\log(m\delta).
\eeq

For the capacity of entanglement we also need the three-point function  \eqref{eq:TTO},
\beq
\langle O KK\rangle =-(2\pi)^2\int_{\Sigma}d^{d-2}y\int_{\Sigma}d^{d-2}{\bar y}\int\limits_{0}^{\infty}dx_1 x_1\int\limits_{-\infty}^0 d{\bar x}_1 {\bar x}_1\int d^d z \langle O(z) T_{00}(x)T_{00}({\bar x})\rangle
\eeq
The appendix in \cite{Rosenhaus:2014zza} provides us with a good guide through the calculation. One identity that will prove useful is
\beq
\int d^{d}x_3 \frac{1}{x_{23}^{\gamma}x_{13}^{\beta}}=\pi^{d/2}\frac{\Gamma(\frac{d-\gamma}{2})\Gamma\left(\frac{d-\beta}{2}\right)\Gamma\left(\frac{\beta+\gamma-d}{2}\right)}{\Gamma(\gamma/2)\Gamma(\beta/2)\Gamma(d-\frac{\beta+\gamma}{2})} x_{12}^{d-\beta-\gamma},
\eeq
where $x_{ij}=x_i-x_j$.

We first compute
\beq
\int_{\Sigma}d^{d-2}y\int_{\Sigma}d^{d-2}{\bar y}\int\limits_{0}^{\infty}dx_1 x_1\int\limits_{-\infty}^0 d{\bar x}_1 {\bar x}_1\int d^d z\frac{(x-z)_{\rho}({\bar x}-z)^{\rho}}{(x-z)^d({\bar x}-z)^d(x-{\bar x})^d}.
\eeq
We can write the integrand in the form
\beq
\frac{(x-z)_{\rho}({\bar x}-z)^{\rho}}{(x-z)^d({\bar x}-z)^d(x-{\bar x})^d}=\frac{1}{(x-{\bar x})^{d}}\frac{\partial_{\rho,x}\partial^{\rho}_{\bar x}}{(d-2)^2}\left(\frac{1}{(z-x)^{d-2}(z-{\bar x})^{d-2}}\right),
\eeq
which we can easily integrate over $z$ to get
\beq
\frac{4 \pi^{d/2}}{\Gamma(\frac{d-2}{2})(d-2)^2}\frac{1}{(x-{\bar x})^{2d-2}}.
\eeq
The remaining integrals can be computed as before. Thus, the contribution from the first term in \eqref{eq:TTO} sums up to
\bear
&&\frac{2 \pi ^{1-\frac{d}{2}} \Gamma \left(\frac{d}{2}\right)^3}{(d-4) (d-2)^2 \Gamma (d)}\mathcal{A}_{\Sigma}\left(\delta^{4-d}-m^{d-4}\right)\\
&&\stackrel{d=4}{=}-\frac{\mathcal{A}_{\Sigma}}{12 \pi}\log(m\delta).
\eear

The contribution from the second term in \eqref{eq:TTO} is more involved. We rewrite the integrand
\bear
&&\frac{(z-{\bar x})_{\rho}(x-{\bar x})^{\rho}(z-x)_{\lambda}(x-{\bar x})^{\lambda}}{(z-x)^d(z-{\bar x})^d(x-{\bar x})^{d+2}} \\
&=& \frac{(x-{\bar x})^{\rho}(x-{\bar x})^{\lambda}}{(x-{\bar x})^{d+2}}\frac{\partial_{\rho,x}\partial_{\lambda,{\bar x}}}{(d-2)^2}\left(\frac{1}{(z-{\bar x})^{d-2}(z- x)^{d-2}}\right),
\eear
which we integrate over $z$ to get
\beq
-\frac{2\pi^{d/2}(d-3)}{\Gamma(\frac{d-2}{2})(d-2)^2}\frac{1}{(x-{\bar x})^{2d-2}},
\eeq
which we know how to integrate with respect to the two other coordinates. The contribution from the second term sums up to
\bear
&&\frac{(d-3) d \pi ^{1-\frac{d}{2}} \Gamma \left(\frac{d}{2}\right)^3}{(d-4) (d-2)^2 \Gamma (d)}\mathcal{A}_{\Sigma}\left(\delta^{4-d}-m^{d-4}\right)\\
&&\stackrel{d=4}{=}-\frac{\mathcal{A}_{\Sigma}}{6\pi}\log(m\delta).
\eear
This gives the contribution of the three point function as
\bear
\langle K\,K \mathcal{O}\rangle &=&\frac{(d-1) \pi ^{1-\frac{d}{2}} \Gamma \left(\frac{d}{2}\right)^3}{\left(d^2-6 d+8\right) \Gamma (d)}\mathcal{A}_{\Sigma}\left(\delta^{4-d}-m^{d-4}\right)\\
&=& -\frac{\mathcal{A}_{\Sigma}}{4}\log(m\delta)
\eear

Summing all the contributions we get
\bear
\langle K\,K\,\mathcal{O}\rangle-2\langle K \mathcal{O}\rangle &=&\mathcal{A}_{\Sigma}\frac{ (d-3) (d-2) \pi ^{\frac{3}{2}-\frac{d}{2}} \Gamma \left(\frac{d}{2}-1\right)^2}{2^{d+1}(d-4) \Gamma \left(\frac{d+1}{2}\right)}\left(\delta^{4-d}-m^{d-4}\right)\\
&\stackrel{d=4}{=}&-\frac{\mathcal{A}_{\Sigma}}{12 \pi}\log(\delta m)
\eear
so in $d=4$ the first order correction to the capacity of entanglement $\Delta S^2_{EE}$ is
\beq
\delta \Delta S^2_{EE} \stackrel{d=4}{=} \frac{m^2}{24\pi}\mathcal{A}_{\Sigma}\log(m\delta).
\eeq
Thus,  under the mass deformation the correction to the capacity of entanglement grows as fast as the correction to the entanglement entropy.

\subsection{A general CFT}\label{gencft}
The three-point function does not vanish in general and is quite complex \cite{Osborn:1993cr}:
\beq
\langle T_{\mu\nu}(x_1) T_{\alpha\beta}(x_2) \mathcal{O}(x_3) \rangle = \frac{I_{\mu\nu,\lambda\kappa}(x_{13})I_{\alpha\beta,\sigma\rho}(x_{23})t_{\lambda\kappa\sigma\rho}(X_{12})}{(x_{12})^{2d-\Delta}(x_{13})^{\Delta}(x_{23})^{\Delta}},
\eeq
where
\bear
t_{\lambda\kappa\sigma\rho} &=& a_1 h^1_{\lambda\kappa}({\hat X})h^1_{\sigma\rho}(\hat{X})+ a_2 h^2_{\lambda\kappa\sigma\rho}({\hat X})+a_3h^3_{\lambda\kappa\sigma\rho},\\
h^1_{\alpha\beta}({\hat X}) &=& {\hat X}_{\alpha}{\hat X}_{\beta}-\frac{1}{d}\delta_{\alpha\beta}\\
h^2_{\lambda\kappa\sigma\rho}({\hat X}) &=& {\hat X}_{\lambda}{\hat X}_{\sigma}\delta_{\kappa\rho}+(\lambda\leftrightarrow\kappa,\sigma\leftrightarrow\rho)\\
&&-\frac{4}{d} {\hat X}_{\lambda}{\hat X}_{\kappa}\delta_{\sigma\rho}-\frac{4}{d} {\hat X}_{\sigma}{\hat X}_{\rho}\delta_{\lambda\kappa}+\frac{4}{d^2}\delta_{\alpha\beta}\delta_{\sigma\rho},\\
h^3_{\lambda\kappa\sigma\rho} &=& \delta_{\lambda\sigma}\delta_{\kappa\rho}+ \delta_{\lambda\rho}\delta_{\kappa\sigma} -\frac{2}{d}\delta_{\lambda\kappa}\delta_{\sigma\rho},\\
I_{\mu\nu,\lambda\kappa}(x) &=& \frac{1}{2}(I_{\mu\lambda}(x)I_{\nu\kappa}(x) +I_{\mu\kappa}(x)I_{\nu\lambda}(x) )-\frac{1}{d}\delta_{\mu\nu}\delta_{\lambda\kappa},\\
I_{\alpha\beta}(x)&=&\delta_{\alpha\beta}-2\frac{x_{\alpha}x_{\beta}}{x^2},\\
\hat{X}_{\mu}&=& \frac{X_{\mu}}{\sqrt{X^2}},\\
X_{12} &=& -X_{21}=\frac{x_{13}}{x_{13}^2}-\frac{x_{23}}{x_{23}^2}.
\eear

The condition of conservation of momentum and energy lead to two conditions for the coefficients $a_1$, $a_2$ and $a_3$,
\bear
a_1+4a_2-\frac{1}{2}(d-\Delta)(d-1)(a_1+4a_2)-d\Delta a_2&=&0\\
a_1+4a_2+d(d-\Delta)a_2+d(2d-\Delta)a_3&=&0.
\eear
Thus, the three-point function is defined up to a multiplicative constant.
Next, we use a few helpful relations \cite{Osborn:1993cr}:
\bear
&&I_{\mu\alpha}(x_1-x_3)X_{12,\alpha} = -\frac{(x_1-x_2)^2}{(x_3-x_2)^2}X_{23,\mu}, \quad I_{\mu\alpha}(x_2-x_3)X_{12,\alpha}
=-\frac{(x_1-x_2)^2}{(x_1-x_3)^2}X_{31,\mu}, \nonumber \\
&&I_{\mu\alpha}(x_1-x_3)I_{\alpha\nu}(x_3-x_2) = I_{\mu\nu}(x_1-x_2)+2(x_1-x_2)^2 X_{23,\mu}X_{31,\nu}.
\eear

The $a_1$ dependent term becomes (sans the factors of $|x_i-x_j|$ in the denominator)
\beq
h^1_{\mu\nu}({\hat X}_{23})h^1_{\alpha\beta}({\hat X}_{31}).
\eeq
The $a_2$ dependent term becomes (the permutations produce 3 additional terms)
\bear
&&\!\!\!\!\!\!\!(\Xhat_{23,\mu}\Xhat_{31,\alpha}(I_{\nu\beta}(x_{12})+2\Xhat_{23,\beta}\Xhat_{31,\nu})+(\mu\leftrightarrow\nu,\, \alpha\leftrightarrow\beta))\nonumber\\
&& +\frac{4}{d}\left(\frac{\delta_{\mu\nu}\delta_{\alpha\beta}}{d}-\Xhat_{23,\mu}\Xhat_{23,\nu}\delta_{\alpha\beta}-\Xhat_{31,\alpha}\Xhat_{31,\beta}\delta_{\mu\nu}\right).
\eear
The $a_3$ dependent term becomes
\bear
&&I_{\mu\alpha}(x_{12})I_{\nu\beta}(x_{12})+I_{\mu\beta}(x_{12})I_{\nu\alpha}(x_{12})+8 \Xhat_{13,\alpha}\Xhat_{13,\beta}\Xhat_{23,\alpha}\Xhat_{23,\beta}\\
&&+2(I_{\mu\alpha}(x_{12})\Xhat_{31,\beta}\Xhat_{23,\nu}+(\mu\leftrightarrow\nu,\, \alpha\leftrightarrow\beta))-\frac{2}{d}\delta_{\mu\nu}\delta_{\alpha\beta}.\nonumber
\eear

The integrals can be computed in a similar manner as before and are straightforward. In this paper, we are only interested in the case where all spacetime indices are set to $0$ and $x_{1,0}=x_{2,0}$.

After integrating over $x_3$, we have the intermediate result
\bea
&&g\!\int_{\mathbb{R}^d}\!\!\!\!{\rm d}^d x_3\langle KKO(x_3)\rangle\\ 
&&\!\!\!\!\!\!= -ga_3(2\pi)^2\frac{\pi^{d/2}f(d,\Delta)\Gamma\left(\frac{d-\eta}{2}\right)^2\Gamma\left(\eta -\frac{d}{2}\right)}{\Gamma \left(\frac{\eta}{2}+2\right)^2\Gamma(d-\eta)}\int\limits_{0}^{\infty}\!{\rm d}x_1\!\int_{\Sigma}\!{\rm d}^{d-2}y\!\!\int\limits_{-\infty}^{0}\!\!{\rm d}{\bar x}_1\!\int_{\Sigma}\!{\rm d}^{d-2}{\bar y}\frac{x_1{\bar x}_1}{(x-{\bar x})^{d+\Delta}}\nonumber
\eea
where
\beq
f(d,\Delta) = \frac{(d-2)(\Delta-1)\Delta(\Delta+1)(\Delta+2)(d-\Delta)(d-\Delta +2)}{8\left((d-1)\Delta^2-2(d-1)d\Delta+(d-2)d(d+1)\right)}.\label{eq:ffunc}
\eeq

\section{More detailed computations of the three-point function integral with spherical entangling surface}\label{sphereintegral}
In this appendix, we try to give more details on the computation of \eqref{eq:xintegral} for the values appearing in the text.

We make use of the Fourier transformation identity
\beq
\frac{1}{x^{d+\Delta}}=\frac{\Gamma\left(\frac{-\Delta-1}{2}\right)}{2^{d+\Delta}\sqrt{\pi}^{d-1}\Gamma\left(\frac{d+\Delta}{2}\right)}\int d^{d-1}k \frac{{\rm e}^{-i {\vec k}\cdot{\vec x}}}{k^{-\Delta-1}}
\eeq
to separate the two integrals over the balls. The integrals over the angular coordinates of $x_1$ and $x_2$ contribute
\beq
\Omega_{d-3}\int\limits_{0}^{\pi}{\rm d}\theta\sin^{d-3}\theta {\rm e}^{i k r\cos\theta}=\Omega_{d-3}\sqrt{\pi}\left(\frac{2}{k r}\right)^{\frac{d-3}{2}}\Gamma\left(\frac{d}{2}-1\right)J_{\frac{d-3}{2}}(k r).
\eeq

After also integrating over the angular coordinates of $k$, we have
\bear\label{complicated}
&&\!\!\!\!\Omega_{d-3}^2\Omega_{d-2}\frac{\pi^{\frac{3-d}{2}}\Gamma\left(\frac{d}{2}-1\right)^2 \Gamma \left(\frac{-\Delta-1}{2} \right)}{2^{\Delta+5}R^{2}\Gamma \left(\frac{d+\Delta}{2}\right)}\!\int\limits_{0}^{\infty}\!\! {\rm d}k \int\limits_{0}^{R}\!\! {\rm d}r_1\!\int\limits_{0}^{R}\!\! {\rm d}r_2\!\frac{\left(R^2-r_1^2\right) \left(R^2-r_2^2\right) J_{\frac{d-3}{2}}(k r_1) J_{\frac{d-3}{2}}(k r_2)}{k^{-\Delta -2}(r_1r_2)^{\frac{1-d}{2}}}\nonumber\\
&&\!\!\!\!=\Omega_{d-3}^2\Omega_{d-2}\frac{\pi^{\frac{3-d}{2}}\Gamma\left(\frac{d}{2}-1\right)^2 \Gamma \left(\frac{-\Delta-1}{2} \right)}{2^{\Delta+3}R^{\Delta-d}\Gamma \left(\frac{d+\Delta}{2}\right)}\!\int\limits_{0}^{\infty}\!\! {\rm d}k \frac{\left(J_{\frac{d+1}{2}}(k)\right)^2}{k^{\Delta+2}},\nonumber
\eear
where we have removed the $R$ dependence from the integral. We can compute the remaining integral explicitly,
\beq
I_{d,\Delta} = \int\limits_{0}^{\infty}\!\! {\rm d}k \frac{\left(J_{\frac{d+1}{2}}(k)\right)^2}{k^{\Delta+2}}= \frac{\pi ^{d-\frac{3}{2}} 2^{3-\Delta } \Gamma \left(1-\frac{\Delta }{2}\right) R^{d-\Delta }}{\left(\Delta ^2-1\right) \Gamma \left(\frac{d-1}{2}\right) \Gamma \left(\frac{1}{2} (d-\Delta +4)\right)}
\eeq
It is still divergent for even values of $\Delta$ but is now finite for odd values of $\Delta$.
The divergences with even $\Delta$ emerge from the UV limit $k\to \infty$, which we will regulate with $k\to \frac{1}{\epsilon}$. The logarithmically divergent term is
\beq
I^{(log)}_{d, even\Delta}= -\frac{\pi ^{d-\frac{5}{2}} 2^{-\Delta } \Gamma \left(-\frac{\Delta }{2}-\frac{1}{2}\right) \Gamma \left(2 \Delta
   -\frac{1}{2}\right)  R^{d-\Delta }}{\Gamma \left(\frac{d-1}{2}\right) \Gamma (2 \Delta ) \Gamma
   \left(\frac{1}{2} (d-\Delta +4)\right)}\log (\epsilon ) \ .
\eeq

\setcounter{equation}{0}

\end{document}